\documentclass[conference]{IEEEtran}
\IEEEoverridecommandlockouts
\usepackage{cite}
\usepackage{amsmath,amssymb,amsfonts}
\usepackage{amsthm}
\usepackage{graphicx}
\usepackage{textcomp}
\usepackage{xcolor}
\usepackage{booktabs}
\usepackage{algorithm}
\usepackage{algpseudocode}
\usepackage{svg}
\usepackage{graphicx}
\usepackage{enumitem}
\usepackage{titlesec}

\newtheorem{proposition}{Proposition}
\newtheorem{lemma}{Lemma}
\newtheorem{theorem}{Theorem}
\newtheorem{definition}{Definition}

\DeclareMathOperator*{\argmax}{arg\,max}

\def\BibTeX{{\rm B\kern-.05em{\sc i\kern-.025em b}\kern-.08em
    T\kern-.1667em\lower.7ex\hbox{E}\kern-.125emX}}

\begin{document}

\IEEEaftertitletext{\vspace{-1\baselineskip}}
\setlength{\textfloatsep}{-1pt}
\setlength{\abovecaptionskip}{-1.6pt}
\setlength{\itemindent}{10cm}

\setlength{\abovedisplayskip}{5pt}  
\setlength{\belowdisplayskip}{5pt}  
\setlength{\abovedisplayshortskip}{5pt}  
\setlength{\belowdisplayshortskip}{5pt}

\title{\LARGE Toward Resilient Airdrop Mechanisms: Empirical Measurement of Hunter Profits and Airdrop Game Theory Modeling
}


\author{\IEEEauthorblockN{Junliang Luo\IEEEauthorrefmark{1},
Hong Kang\IEEEauthorrefmark{2},
Shuhao Zheng\IEEEauthorrefmark{3},
Xue Liu\IEEEauthorrefmark{4}}
\IEEEauthorblockA{School of Computer Science, McGill University, Montréal, Québec, Canada\\
\IEEEauthorrefmark{1}junliang.luo@mail.mcgill.ca,
\IEEEauthorrefmark{2}hong.kang@mail.mcgill.ca,
\IEEEauthorrefmark{3}shuhao.zheng@mail.mcgill.ca,
\IEEEauthorrefmark{4}xueliu@cs.mcgill.ca}
}

\maketitle

\begin{abstract}
Airdrops issued by platforms are to distribute tokens, drive user adoption, and promote decentralized services.
The distributions attract airdrop hunters (attackers), who exploit the system by employing Sybil attacks, i.e., using multiple identities to manipulate token allocations to meet eligibility criteria.
While debates around airdrop hunting question the potential benefits to the ecosystem, exploitative behaviors like Sybil attacks clearly undermine the system's integrity, eroding trust and credibility.
Despite the increasing prevalence of these tactics, a gap persists in the literature regarding systematic modeling of airdrop hunters' costs and returns, alongside the theoretical models capturing the interactions among all roles for airdrop mechanism design.
Our study first conducts an empirical analysis of transaction data from the Hop Protocol and LayerZero, identifying prevalent attack patterns and estimating hunters' expected profits. Furthermore, we develop a game-theory model that simulates the interactions between attackers, organizers, and bounty hunters, proposing optimal incentive structures that enhance detection while minimizing organizational costs.

\end{abstract}

\vspace{-0.03cm}


\section{Introduction}
Decentralized Finance (DeFi) uses cryptocurrency tokens to offer global financial services without relying on traditional banks \cite{chang2020blockchain}.
To expand their reach and incentivize engagement, DeFi service providers often conduct airdrops \cite{allen2023airdrop}, a strategy to distribute free tokens to eligible users' wallets to promote adoption, reward community members, and achieve a decentralized token distribution \cite{investopedia_airdrop}.
Airdrops can be distributed to early adopters of a service, such as users of token exchanges, active participants in NFT-focused social media, or players in GameFi’s play-to-earn, where tokens are awarded directly by engaging in gameplay with on-chain assets \cite{proelss2023gamefi}.
Regarding potential benefits, service users and community members gain advantages from airdrops, while airdrop hunters are similarly attracted to exploit arbitrage opportunities \cite{yaish2024tierdrop}.
Irrespective of the role of the participants i.e., legitimate users or hunters, receiving an airdrop requires engaging in the activities that align with the eligibility criteria set forth by the platforms.
Previous studies that categorized the participant roles emphasized that the hunters act strategically through broad farming to economically gain from the airdrop distribution,  are not considered genuine users since mercenary users will likely not be the future users of the protocol \cite{allen2023crypto}.
In this context, farming refers to behaviors of actively participating in required actions of criteria to qualify for airdrops.
While farming (strategically participating in airdrops to earn token rewards) through standard participation in airdrops may be considered legitimate and whether the farmers benefit the ecosystem is contentious \cite{yaish2024tierdrop}, employing deceptive tactics such as a single entity using multiple identities or wallets to manipulate token distribution (a practice known as a Sybil attack) is generally viewed as unethical and harmful to token prices volatility \cite{bhambhwani2023governing}, and erosion of trust and credibility \cite{tonresearch2024sybil}.
The ethical implications and potential disruptions caused by such practices necessitate closer scrutiny of airdrop hunters to maintain the integrity of airdrops.
Currently, a notable lack persists within the existing literature in systematically quantifying the extent of the behaviours of airdrop hunters, and evaluating the associated costs and returns in past airdrops.
Additionally, a gap exists in the development of a game theory model to guide platforms in optimizing airdrop incentive mechanisms while minimizing organizational costs and deterring exploitative behaviors.

\vspace{-0.1em}
The economic disruptions and potential erosion of trust caused by the actions of airdrop hunters necessitate the study of measurement and mechanisms to characterize their behaviors and interactions for the sustainability of airdrop systems.
To progress towards a framework improving airdrops mechanism design, it is imperative to identify and measure hunter patterns and quantifying the magnitude of associated costs and returns through exploitative behaviors. 
Subsequently, research should be conducted on designing airdrop mechanisms that incorporate incentivized reporting (where users report hunters for additional rewards) and self-reporting (allowing hunters to return extracted funds while retaining a small portion), modeled through a theoretical framework. 
The theoretical framework could guide in developing future equitable airdrop practices that protect genuine participants, proceeding us toward the future of airdrop system sustainability rather than opportunistic exploitation or escapism.

\vspace{-0.1em}
In the study, we begin with an empirical analysis of transaction data from reported airdrop attackers on the Hop Protocol and LayerZero, with a systematic evaluation of behavior patterns and the cost-expected profits of airdrop hunters. Further, we propose a game-theory model to capture the interactions between organizers, attackers, and bounty hunters, modeling a reward policy that incentivizes detection while minimizing organizer costs.
The contribution is stated to be:

\vspace{0.05em}
\begin{enumerate}[itemsep=2pt,topsep=-1pt,parsep=0pt,leftmargin=11pt] 
    \item We collect and analyze two airdrop datasets from Hop Protocol and LayerZero, defining distinct transaction patterns and behaviors indicative of airdrop hunter behaviors. We systematically categorize airdrop hunter behaviors including funding and common receivers, sequential bridging, uniformity behaviors, and measure the prevalence of these patterns in both datasets.
    
    \item We conduct an expected profit modeling analysis of Hop's reported hunters to estimate their potential rewards, even in reality hunters were disqualified and receiving no actual payouts. The findings reveal that while many hunter groups could achieve positive net profits, others were economically unfeasible, providing insights into cost-profit details. 
    
    \item We propose a game theory model to simulate the interactions between hunters, bounty hunters (reporters seeking rewards), organizers, and reporting mechanisms, providing a framework for optimal rewards for incentivizing reporting and detection strategies while minimizing costs for the organizers. 
\end{enumerate}

\section{Related Work}
Research most relevant to our study includes research on the financial and social roles and effects of airdrops \cite{makridis2023rise,bauer2019airdrops}, airdrops privacy concerns \cite{harrigan2018airdrops}, and studies focused on detecting or avoiding sybil attacks in airdrops \cite{allen2023crypto,liu2022fighting,ohlhaver2022decentralized}.
Some works targeted the issue of detecting malicious accounts in airdrops extend beyond officially reported attackers by using clustering algorithms to identify suspicious behavior \cite{liu2022fighting}.
However, none of the existing methods systematically modeled the cost and return dynamics of the attack behaviors, which is essential for designing more effective defenses and optimizing reward distribution mechanisms.
In addition to studies on detecting attacks, several works examined airdrop design strategies.
Allen \cite{allen2023crypto} explored airdrops as a coevolutionary process between protocols and recipients, where both parties adapt their strategies over time, focusing on design innovations such as task-based claiming and multi-round airdrops to optimize distribution and minimize mercenary behavior.
Yaish et al. \cite{yaish2024tierdrop} explored the role of airdrop farmers, building on previous work to argue that instead of solely preventing farming behavior, platforms can harness it to strengthen network effects and drive user growth.

\section{Data Collection}
\noindent \textbf{Hop Reported Attackers Transactions}.
We collected a dataset based on reported attackers from the Hop Airdrop GitHub repository\footnote{https://github.com/hop-protocol/hop-airdrop},  drawing from 150 validated reported groups corresponding to user-reported hunters (attackers) identified through closed GitHub issues, including 3,551 unique wallet addresses. 
The repository contains a record of transfer only noting unique pairs of \textit{from} and \textit{to} between these attackers and some externally owned accounts (EOAs), lacking details on the exact multiple native transactions, ERC20 token transfers, and their token types, token volumes, and timestamps, gas fee, gas used between these pairs. 
Therefore, we conducted an extraction of all native transactions and ERC20 token transfers across the five blockchains: Ethereum, Optimism, Arbitrum, Polygon, and xDai, supported by Hop Protocol back in 2022 before the airdrop, specifically filtering for transactions involving the reported attacker addresses and their associated EOAs in interaction with Hop Protocol contracts.

\noindent \textbf{LayerZero Reported Attackers Transactions}.
LayerZero provided a list of reported attackers on the GitHub repository\footnote{https://github.com/LayerZero-Labs/sybil-report}, along with all transactions made prior to the snapshot for the native token airdrop on May 1, 2024.
The dataset includes the source and destination chains, transaction hashes, sender wallets, timestamps, and the monetary value in USD.
We excluded addresses from the initial lists provided, retaining only the reported attackers from 198 groups of 7,681 addresses, and the transactions associated with these reported attackers.

\begin{table}[htbp!]
\centering
\caption{Extracted transactions and token transfers}
\begin{tabular}{@{}lrr@{}}
\toprule
\textbf{Metric}                & \textbf{Hop Protocol} & \textbf{LayerZero} \\ 
\midrule
Attacker addresses             & 3,551                 & 7,681              \\
Transactions                   & 3,590,322             & 127,339,267        \\
Filtered transactions          & 29,936                & 995,374            \\
ERC20 token transfers          & 4,387,230             &                \\
Filtered ERC20 token transfers & 173,978               &                \\
\bottomrule
\end{tabular}
\label{data_collected}
\end{table}
\vspace{-0.57em}

\section{Airdrop Hunter Behavior Patterns}
To systematically analyze attacker behaviors, we generalize the common patterns observed across both datasets of customer-reported attackers.
Drawing on these observations, we abstract the behaviors into the following patterns: initial funding, cross-chain transfers, and sequential bridging, commonly recognized by airdrop organizers as strong indicators for validating reported attackers.

\label{farming_pattern}
\noindent \textbf{\textit{Initial funding and common receiver}}:
Consider a set of addresses \( A = \{A_i\}_{i=1}^n \) funded by a limited set of source addresses \( S = \{A_{s_j}\}_{j=1}^k \), with \( k \ll n \). Additionally, a set of common receivers \( R = \{A_{r_j}\}_{j=1}^m \), where \( m \ll n \), consolidates transactions from multiple attacker addresses \( A_i \) across chains into a few addresses.
Let $T(A_i, A_j)$ represent a transaction from address $A_i$ to address $A_j$. 
\begin{itemize}[label={}, leftmargin=10pt, itemsep=8pt, topsep=0pt, parsep=0pt]
    \item $\mathit{Same\text{-}Chain\ Funder}:$ The initial funding transactions from a source address $S_j$ to a target address $A_i$ on the same chain, occur as follows:
    \[
    \forall A_i \in A, \quad \exists A_{s_j} \in S \; \text{, such that} \; T(A_{s_j}^{C_0}, A_i^{C_0}).
    \]

    \item $\mathit{Cross\text{-}Chain\ Receiver}:$ The common receiver $A_r$ collects transactions from various addresses $A_i$ across different chains $\{C_{k}\}_{k\in K}$, which are consolidated into one address on a single destination chain $C_1$:
    \[
    \forall A_i \in A, \quad \exists A_{r_j} \in R \; \text{, such that} \; T(A_i^{C_k}, A_{r_j}^{C_1}).
    \]
\end{itemize}

These initial funding transactions may or may not directly contribute to airdrop farming, i.e., purposefully interacting with a protocol to meet the eligibility criteria for receiving airdrops of tokens distributed by the protocol.
The initial funding could be used by attackers to conduct a test to ensure that the address is set up and capable of receiving  transactions, or to hold the bare minimum amount of tokens needed to start operations, including transferring bridge assets or interacting with other smart contracts.
\noindent \textbf{\textit{Sequential transfer}}: It describes a transaction chain of sequential bridging flows.
The transactions occur between any pair of addresses in the set $A$ across different chains.
Each address $A_i$ may exist on multiple chains. $A_i^{C_p}$ denotes address $A_i$ on chain $C_p$.
The characteristics include:
\begin{itemize}[label={}, leftmargin=10pt, itemsep=8pt, topsep=0pt, parsep=0pt]
    \item $\mathit{Flow\ of\ Funds:}$ Transactions maintain a close value flow:
    \[
    |F(T_{i+1}) - F(T_i) + \delta_i| < \epsilon, \quad \forall i \in [1, m-1].
    \]

    \item $\mathit{Cross\text{-}Chain\ Requirement:}$ Transactions span multiple chains (through interacting with a bridge protocol):
    \[
    \forall T_i, T_{i+1} \; \text{such that} \; T_i(A_j^{C_p}, A_k^{C_q}), \; \text{with} \; C_p \neq C_q.
    \]

    \item $\mathit{Activity\ Time\ Window:}$ Transactions occur within a specific time window $\Delta t$:
    \[
    |t(T_i) - t(T_{i+1})| < \Delta t, \quad \forall i \in [1, m-1].
    \]
\end{itemize}

where $F(T_i)$ is the amount of funds transferred in transaction $T_i$, $\delta_i$ is the transaction fee for $T_i$, which varies depending on network conditions, $m$ represents the total number of  transactions in the observed flow of funds, and $\epsilon$ represents the allowable variation in transferred amounts.
\noindent \textbf{\textit{Uniformity}}: For a subset of addresses in $A$: $A' = \{A_i\}_{i=1}^k$, the activity pattern is characterized by uniform bridging transaction counts and consistent transaction volumes across these addresses. This suggests coordinated behavior intended for optimizing airdrop eligibility. 
The uniformity in transaction count:
$$
\forall A_i, A_j \in A', \quad \left| \, \left| T(A_i) \right| - \left| T(A_j) \right| \, \right| < \epsilon_T.
$$
where $\left| T(A_i) \right|$ represents the total number of transactions sent from address $A_i$ to any other addresses in $A$, and $\epsilon_T$ is an integer threshold indicating the allowable difference in transaction counts.
Additionally, the consistency in total transaction volume of tokens is captured by:
$$
\forall A_i, A_j \in A', \quad \left| \sum_{T \in \mathcal{T}(A_i)} F(T) - \sum_{T' \in \mathcal{T}(A_j)} F(T') \right| < \epsilon_F.
$$
where $F(T)$ denotes the volume of transaction $T$, and $\mathcal{T}(A_i)$ is the set of transactions sent from address $A_i$ to any other addresses in $A$. 
The parameter $\epsilon_F$ specifies the allowable variation in total transaction volume between addresses.

\begin{figure}[htbp!]
    \centering
    \resizebox{0.44\columnwidth}{!}{%
    \begin{minipage}[b]{0.441\linewidth}
        \centering
        \includegraphics[width=\linewidth]{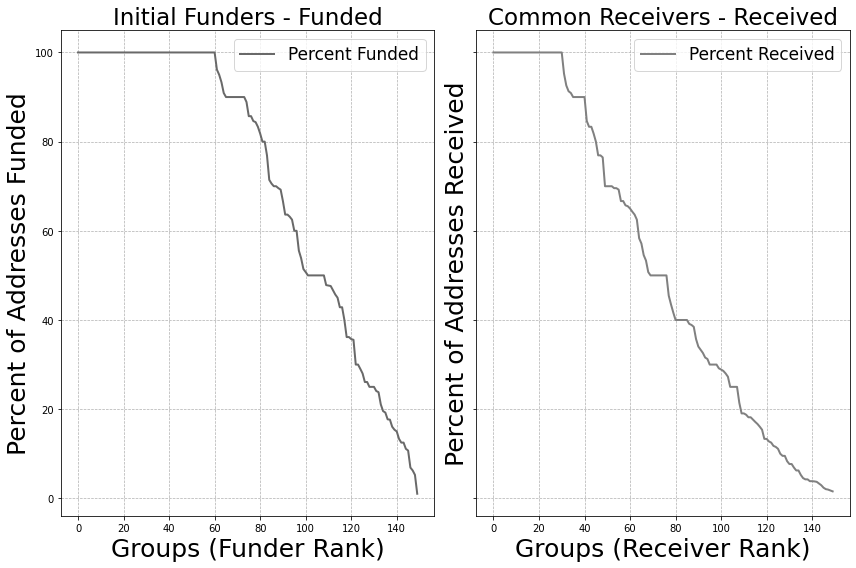}
        \caption{Percent of addresses funded by an initial funder, and sent to a common receiver in each group}
        \label{fig:funder_receiver}
    \end{minipage}
    }
    \hfill
    \resizebox{0.536\columnwidth}{!}{%
    \begin{minipage}[b]{0.544\linewidth}
        \centering
        \includegraphics[width=\linewidth]{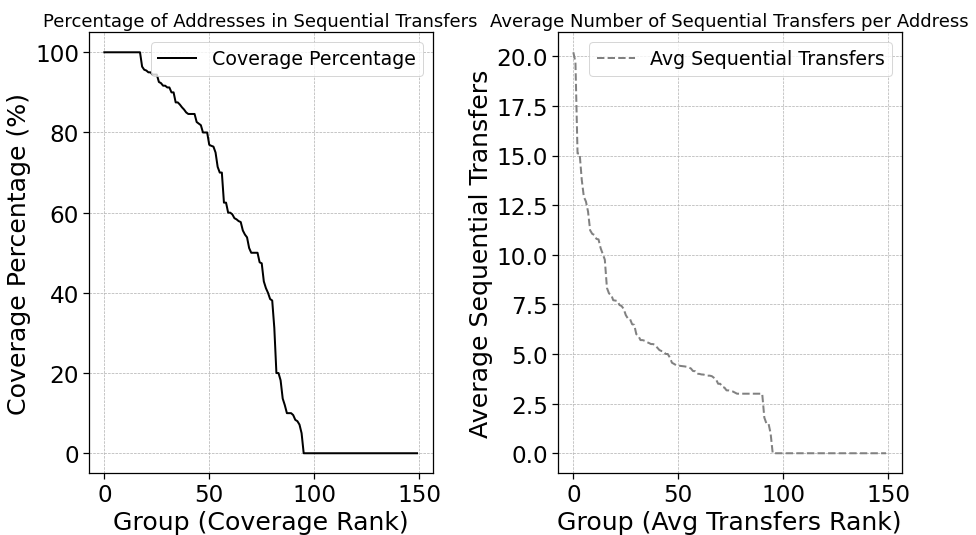}
        \caption{Percent of addresses involved in sequential transfers and the avg number of sequential transfers per address in each group}
        \label{fig:coverage_avg_sequential_transfer_per_address}
    \end{minipage}
    }
\end{figure}

\subsection{Measuring on Hop Protocol}

\noindent \textbf{Initial funding and common receiver}:
Based on the initial funding and final receiving pattern defined in Section \ref{farming_pattern}, we implemented Algorithm \ref{alg:initial_funder_common_receiver_detection} in Appendix \ref{pattern_measure_alg} to quantify both patterns.
We intended to identify the address that transfers funds to the highest number of recipients, and the address that receives transfers from the most other addresses within a group, since these patterns serve as strong evidence of airdrop farming activity likely operated by a single entity.
Also, we visualized percentages of addresses that interacted with the identified address within the group in Figure \ref{fig:funder_receiver}.
The results from applying the detection on the Hop data indicate that 83/46 out of 150 groups exhibit an initial funder/common receiver responsible for funding/receiving from over 80\% of the other addresses within the group, indicating a pronounced centralization (a small number of addresses are responsible for funding or receiving transactions from a large number of other addresses).
And 61/31 out of 150 groups have a single initial funder/common receiver interacts with all other addresses within the group, suggesting the behavior characteristic of single entity schemes of airdrop farming.

\noindent \textbf{Sequential transfer}:
Next phase, we focus on detecting sequential transfers, which along with identifying the initial funder, serve as a clear indicator of coordinated attacker activity. Closely timed and similarly valued transactions often reveal deliberate attempts to create records of bridge usage from the same source of funds.
The detection algorithm is presented in Algorithm \ref{alg:sequential_transfer_detection} (in Appendix \ref{pattern_measure_alg}).
%


\begin{figure}[t!]
    \centering
    \includegraphics[width=0.46\textwidth]{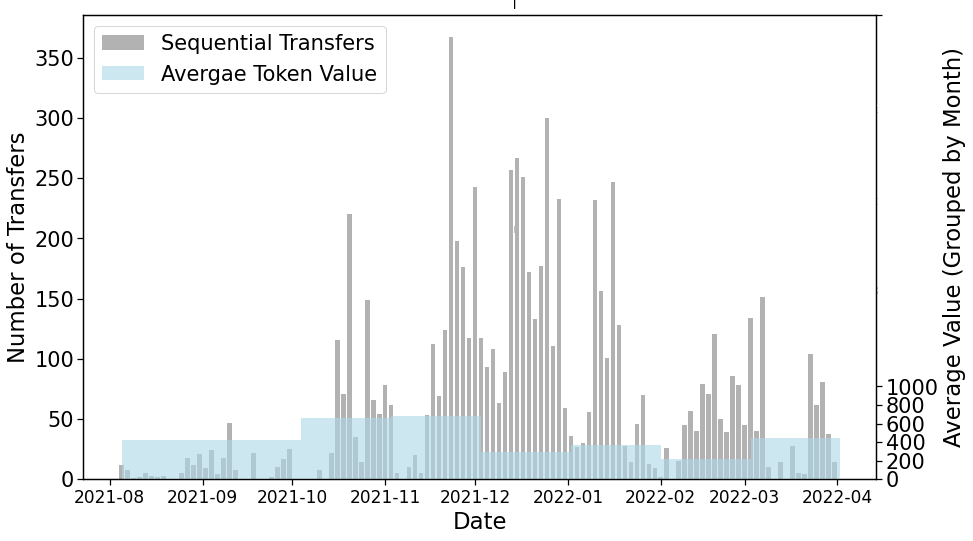}
    \caption{Number of sequential transfers and average token value per transfer over time (grouped by 30 days).}
    \label{fig:sequential_transfer_over_dates}
\end{figure}

We applied the detection algorithm on the Hop dataset with a value threshold $\varepsilon$ of 1\% difference in the token sent value, a time threshold $\Delta_t$ of 30 minutes.
%
%
The number of sequential transfers is concentrated at values below 16. 
The results show that larger groups with more addresses tend to have a higher number of sequential transfers. 
Additionally, the average number of addresses per sequential transfer stays within a small range of 2 to 3.75, and there exists a decreasing trend in the average number of addresses per transfer as the total number of sequential transfers increases. 
This suggests that as sequential transfers increase, fewer addresses are involved in transfers, indicating a potential pattern of diminishing participation per transfer within larger groups.

\begin{figure*}[t!]
    \centering
    \begin{minipage}{0.485\linewidth}
        \centering
        \includegraphics[width=\linewidth]{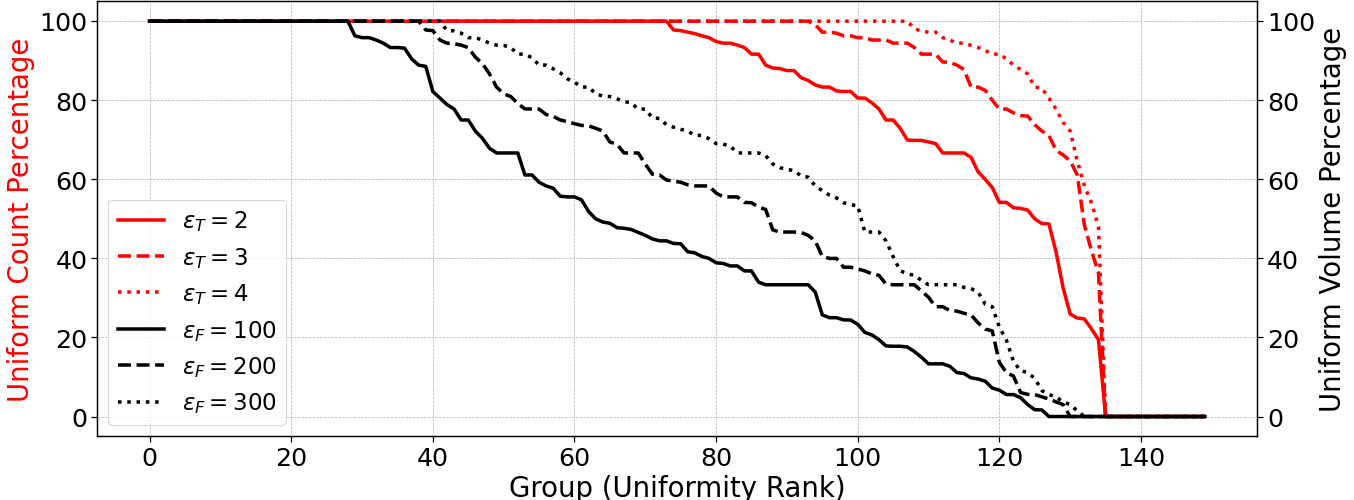}
        \caption{Uniformity of transaction count and volume in the Hop dataset}
        \label{fig:uniformity_volumn_count}
    \end{minipage}
    \hfill
    \begin{minipage}{0.485\linewidth}
        \centering
        \includegraphics[width=\linewidth]{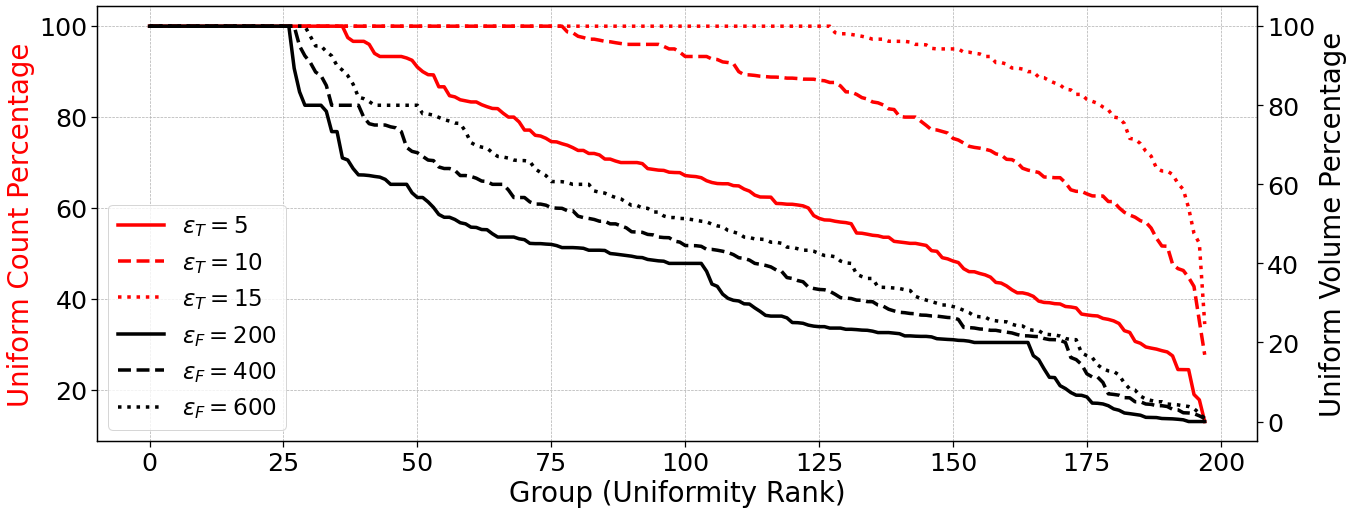}
        \caption{Uniformity of transaction count and volume in the LayerZero dataset}
        \label{fig:uniformity_volumn_count_lz}
    \end{minipage}
\end{figure*}

To assess the extent of address involvement in sequential transfers, we visualize the percentage of addresses within each group that are covered (participating in one or more sequential transfers) in Figure \ref{fig:coverage_avg_sequential_transfer_per_address}. 
The results indicate that half of the groups (74/150) have over half of their addresses involved in sequential transfers.
Additionally, for the covered addresses, we examine the average number of sequential transfers per covered address at the group level as shown in Figure \ref{fig:coverage_avg_sequential_transfer_per_address}. 
The average number of sequential transfers per (covered) address falls between 5 and 15 for the majority of the groups, indicating that a substantial portion of covered addresses are repeatedly involved in multiple sequential transfers.
Figure \ref{fig:sequential_transfer_over_dates} visualizes the number of sequential transfers and the average token value over time. 
Notably, a significant spike in sequential transfers appeared in December 2021 and January 2022, aligning with the period when rumors of upcoming airdrops were widely reported in the press \cite{cryptobriefing2022airdrops}.
The average token value (mostly stablecoin) exhibits a relatively flat trend, indicating that despite the surge in transfer activity, attackers did not engage in high-value transactions, reflecting an attempt to remain inconspicuous but maintain sufficient values in transactions for airdrop eligibility.

\noindent \textbf{Uniformity}: 
To further identify farming patterns, uniformity in both transaction count and transaction volume serves as another marked indicator of coordinated attacker behavior.
Uniformity patterns, where a single entity likely engages in repeated and similar transactions across multiple addresses, are indicative of systematic exploitation. 
Algorithm \ref{alg:uniformity_detection} (in Appendix \ref{pattern_measure_alg}), for uniformity calculation, has been implemented to detect uniform patterns among addresses within a group, using two thresholds of transaction count $\varepsilon_T$ and transaction volume $\varepsilon_F$.
Figure \ref{fig:uniformity_volumn_count} illustrates the uniformity of transaction counts and volumes (number of tokens) across the groups given varying thresholds $\varepsilon_T$ and $\varepsilon_F$. 
As the threshold values increase, both transaction count and volume uniformity percentages rise, meaning that a broader range of transactions are classified as uniform with more lenient thresholds. 
In general, these bridge transactions within each group exhibit high uniformity in transaction volumes and counts across all thresholds, suggesting that attackers maintain consistent behavior in the transaction count and the number of tokens transferred for the majority of addresses within the group.
The figure demonstrates that for the smallest threshold $\varepsilon_T$ of 2, 100 groups (66.6\%) exhibit uniformity in 80\% of transactions in terms of the count.
As $\varepsilon_F$ increases to 300, only about 67 (44.6\%) groups reach 80\% uniformity in transaction volumes among addresses, indicating increased variability in the amounts transferred.
This suggests that while the attackers consistently execute similar numbers of transactions across the addresses, the results demonstrate more fluctuation in the volume of tokens, even under more lenient thresholds.
%


\begin{figure}[t]
    \centering
    \includegraphics[width=0.43\textwidth]{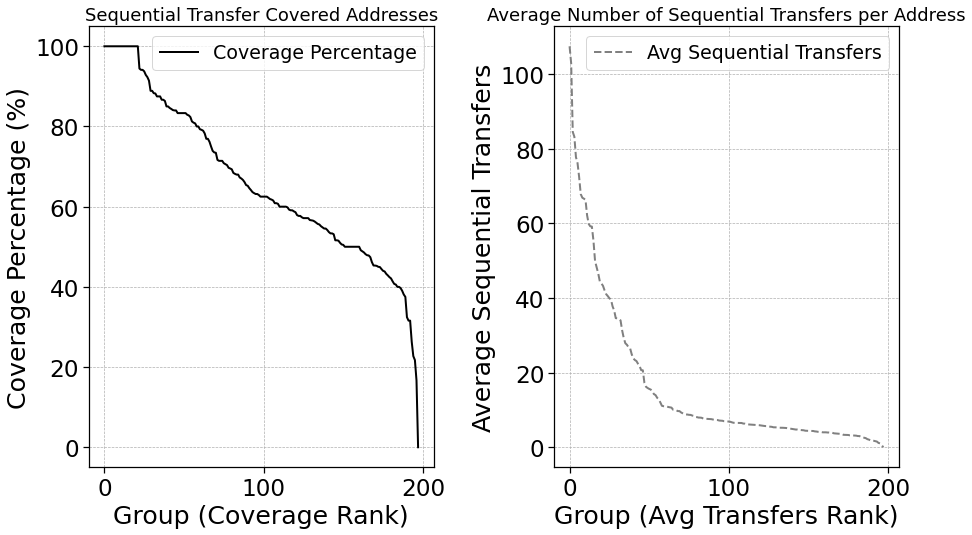}
    \caption{Coverage of sequential transfer addresses and average transfers per address at the group level.}
    \label{fig:coverage_avg_sequential_transfer_per_address_lz}
\end{figure}

\begin{figure}[t]
    \centering
    \includegraphics[width=0.48\textwidth]{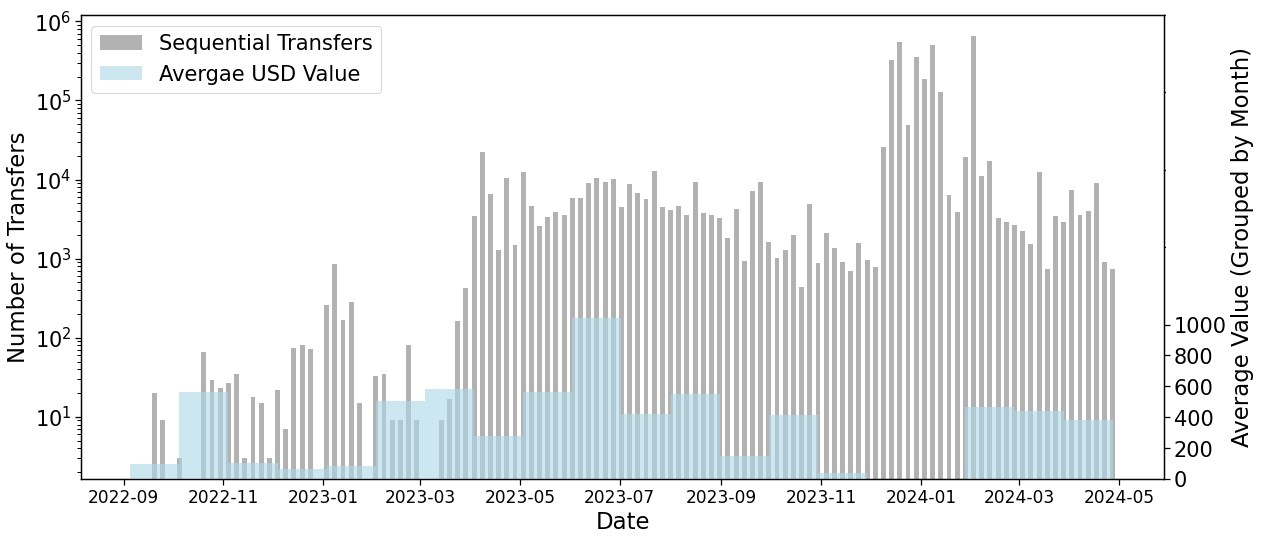}
    \caption{Number of sequential transfers and average USD value per transfer over time (grouped by 30 days).}
    \label{fig:sequential_transfer_over_dates_lz}
\end{figure}

\subsection{Measuring on LayerZero}
We approximate sequential transfers by identifying transactions within-group addresses, transaction values within a margin (1\%), and timestamps occurring within a time window (30-min) to form potential sequential transfers.
This approach offers an approximation due to the absence of true receiver wallet information, where destination transaction hashes reference LayerZero executor or relayer addresses, complicating the ability to definitively trace the recipient.
Future work may include developing methods to identify the true receivers. 
Also, we perform the same uniformity check by evaluating transaction counts and volumes across sender addresses.
Figure \ref{fig:coverage_avg_sequential_transfer_per_address_lz} presents the results indicating that a large proportion of groups (161/198) exhibit more than half of their addresses involved in sequential transfers, suggesting widespread participation in coordinated transfer activities. 
The distribution of average sequential transfers per address shows a steep decline, indicating a heavy-tailed pattern where a small number of groups exhibit higher sequential transfer activity.
Figure \ref{fig:sequential_transfer_over_dates_lz} presents the temporal distribution of sequential transfers alongside the average monetary USD value per transfer grouped by 30-day intervals. 
Following the LayerZero airdrop announcement in Dec 2023 \footnote{https://cryptonews.com/cryptocurrency/layerzero-airdrop/}, there was a spike in the number of sequential transfers.
Despite the volume surge, the average monetary value per transfer during this period is substantially lower than the pre- and post-announcement phases.
Figure~\ref{fig:uniformity_volumn_count_lz} demonstrates the uniformity of transaction count and volume for various thresholds of $\varepsilon_T$ and $\varepsilon_F$. 
For $\varepsilon_T$ of $5$, only one-third of groups achieve $80\%$ uniformity in transaction count, indicating low consistency across addresses. 
As the threshold increases to $10$, this rises to nearly two-thirds of groups reaching the same level of uniformity. 
LayerZero attackers conduct $129$ transactions per address on average, much higher compared to Hop Protocol.
Consequently, the intra-group uniformity in transaction count among attack addresses is also notably high.
However, even at higher values of $\varepsilon_F$, the transaction volumes (USD) remain highly diverse, reflecting a lack of uniformity in the value of transfers. This suggests that while attackers may execute a similar number of transactions across their addresses, the USD amounts transferred vary considerably.

\section{Profit Modeling on Hop Protocol}

We perform a cost-return analysis at the group level for the hunters.
The Hop Protocol facilitates cross-chain asset transfers by bridging users' tokens between Layer 2 networks and Ethereum. 
To qualify for the airdrop as early users, users must have completed at least two bridge transactions with a total transfer volume of \$1,000 or more.
For an address \( i \), the reward \( R_i \) represents the total value of Hop tokens airdropped to this address, based on the protocol's reward distribution rule. Specifically, the reward \( R_i \) is calculated as:
$$
R_i = B \times m_{e,i} \times m_{v,i} \times P,
$$
where $B = 330.4883$ \footnote{https://app.hop.exchange/airdrop/preview} is the constant base amount of tokens by Hop Protocol allocated per address, and $P = 0.1321$ is the Hop token price in USD when the tokens launched on June 09, 2022 \footnote{https://www.coingecko.com/en/coins/hop-protocol}. 
The final reward depends on how early the address interacted with the protocol and the total volume of assets bridged through Hop.
The early bird multiplier $m_{e,i}$ for address $i$ ranges from 2x to 1x, depending on how early the address used the Hop bridge, decreasing linearly over time (The earliest date is June 17, 2021, and the latest is April 1, 2022). 
The volume multiplier $m_{v,i}$ is determined by the bridge value in USD $V_i$, where:
$$
m_{v,i} =
\begin{cases} 
  1, & \text{if } 1,000 \leq V_i < 2,000, \\
  2, & \text{if } 2,000 \leq V_i < 3,000, \\
  3, & \text{if } V_i \geq 3,000.
\end{cases}
$$

The hunters (attackers) are reported in groups, with each group comprising multiple addresses that are likely to be operated by the same individual or entity \footnote{https://github.com/hop-protocol/hop-airdrop/issues}.
To determine the optimal net profit $NP_i$ for each address $i$ in the group, the net profit for address $i$ is given by:
$$
NP_i = R_i - C_i = \left( B \times m_{e,i} \times m_{v,i} \times P \right) - \left( g_i + B_{\phi}(V_i) \right),
$$
where $g_i$ represents the total gas fee (in USD, based on the token price at the transaction time) of all the Hop bridge transactions for address $i$.
The function $B_{\phi}(V_i)$ represents the total bridge fee cost for address $i$, which includes AMM swap fees, slippage, bonder fees, destination chain gas fees, and a minimum fee of $0.25$ \footnote{https://docs.hop.exchange/basics/fees}. 
The fee is complex since its aforementioned components depend on factors including transaction volume, liquidity condition, the specific route of the transfer, and network condition, which vary over time ($\phi$ denotes all conditions affecting the bridge fee).
The hunters theoretically should target maximizing the total net profit for the group, as given by $\sum_{i=1}^{N} \text{NP}_i$.
The hunters seeking to maximize their expected net profit in the Hop airdrop, can optimize their strategy by selecting bridge values $V_i$ just above the thresholds where the volume multiplier $m_{v,i}$ increases specifically at $1,000$, $2,000$, and $3,000$ to maximize their Return on Attack (RoA) relative to cost. 

\noindent \textbf{Measuring the Expected Profit}:
The reported attacker addresses were disqualified from receiving airdrop rewards from Hop Protocol for their bridge transactions as Hop officially excluded them regardless of bridge volume.
Though these addresses still received rewards as liquidity providers, since the liquidity provider airdrop was based solely on the liquidity provision’s amount and duration without needing sybil resistance.
Subsequently, we intend to measure the expected profit for the attackers, since estimating the expected reward allows us to assess the potential profit margin attackers could have gained if undetected, providing concrete evidence of whether the airdrop exploitation was economically motivated.
Therefore, we conducted an experiment to estimate the expected profit for each attacker by analyzing their bridge transactions to calculate the relevant early bird and volume multipliers, and further calculating the expected return as the difference between the reward and the total fees in USD. 
The total bridge fee was calculated by first taking the difference between the token value sent on the source chain and the token value received on the destination chain, which includes all the component bridge fees. 
The differential was then multiplied by the historical daily closing price of the respective token (e.g., Eth, Matic) at the time of the transaction to derive the equivalent USD, and gas fees were similarly calculated using the historical daily closing price of the native token on the transaction date.
We calculated the expected rewards and total fees for each attacker address and reported the results at the group level.

\begin{figure}[t!]
    \centering
    \includegraphics[width=0.9\linewidth]{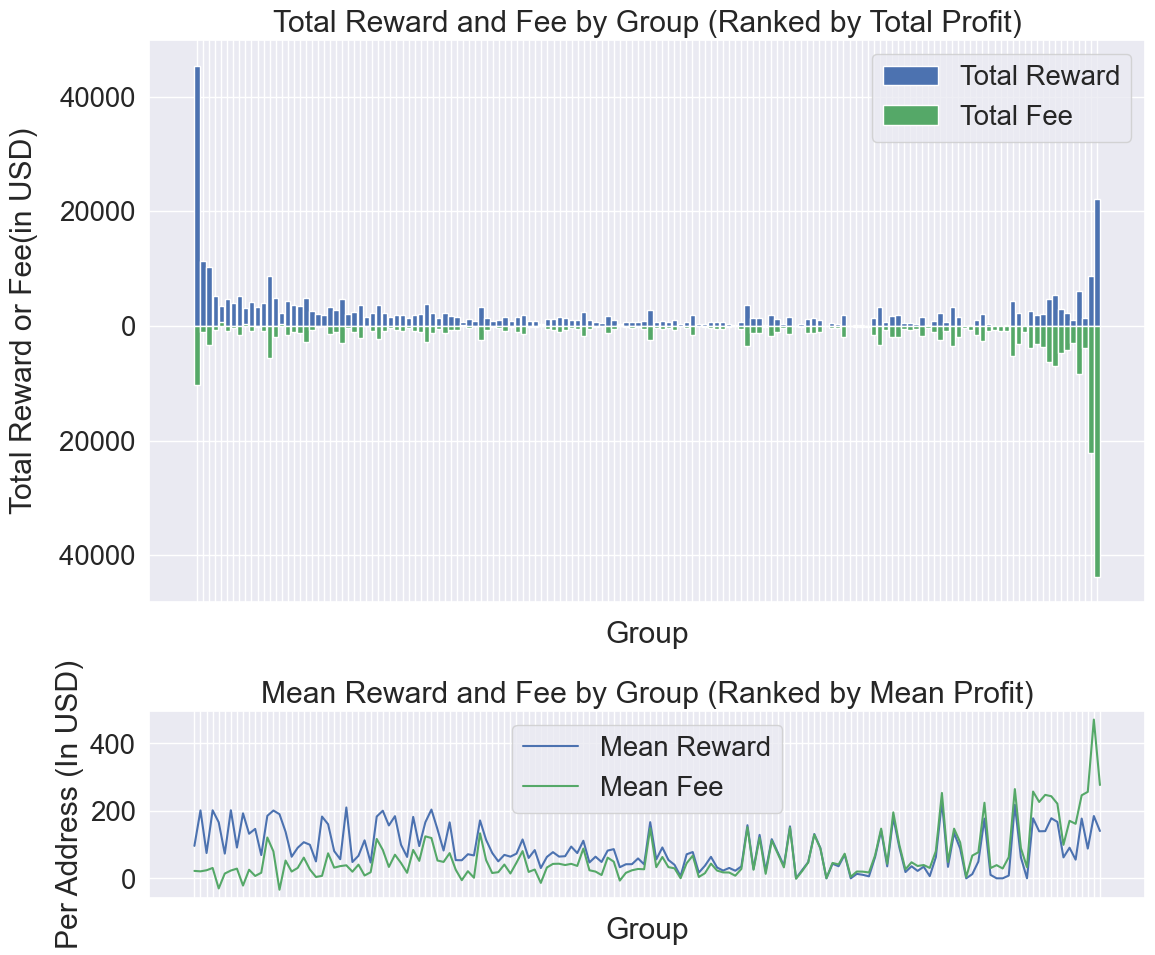}
    \caption{Reward against fee by group ranked by profit.}
    \label{fig:reward_fee_dist}
\end{figure}

\begin{table}[b!]
\centering
\caption{Examples of attacker addresses with the highest expected rewards}
\resizebox{1.0\columnwidth}{!}{%
\begin{tabular}{llcccc}
\hline
\textbf{Group} & \textbf{Address} & \textbf{\begin{tabular}[c]{@{}c@{}}Bridge value\\ USD\end{tabular}} & \textbf{\begin{tabular}[c]{@{}c@{}}Early bird \\ multiplier\end{tabular}} & \textbf{\begin{tabular}[c]{@{}c@{}}Volume \\ multiplier\end{tabular}} & \textbf{\begin{tabular}[c]{@{}c@{}}Reward\\ USD\end{tabular}} \\ \hline
GH issue \#564 \quad\quad\quad & 0x90ff89c637fd1537e151b8ced1d1d0cb94e31ca6 & 14187.228 & 2.6045 & 3 & 341.1181 \\
GH issue \#457 & 0x376110452c2fdae8f842bd1fd4234a2d0452e40c & 5853.667 & 2.5109 & 3 & 328.8595 \\
GH issue \#564 & 0x762d5548ea30bfbf09abd153d0a0ceae75ab418cf & 85397.280 & 2.4967 & 3 & 327.0050 \\
GH issue \#346 & 0xe636c37c8161d1097090534e294cabae78dcdca3 & 10830.814 & 2.2394 & 3 & 293.2999 \\
GH issue \#199 & 0x4125caf9dd93ec6ad7a4c668ec55102f8f18203a & 66427.885 & 2.1671 & 3 & 283.8287 \\
GH issue \#400 & 0x484adfda2b22b76109087bfffcb1d120f8e045a9 & 5545.822 & 2.1164 & 3 & 277.1880 \\
GH issue \#3 & 0xec561dd73346f761d1c09f46999bc8ea1b1e95a2 & 12483.266 & 2.0716 & 3 & 271.3210 \\
GH issue \#239 & 0x47e5d263778b1d6afb4468f92fc7b09b306d7997 & 15207.568 & 2.0572 & 3 & 269.4368 \\
GH issue \#506 & 0xaa6c7b556449e1d5a75b6322a256180edd1c8edc & 5892.124 & 2.0405 & 3 & 267.2485 \\
GH issue \#377 & 0x1ba1ff69f03d38af7ef40e3afdc53ad073eb4642 & 16634.560 & 1.8041 & 3 & 236.2856 \\ \hline
\end{tabular}
}
\label{tab:address_high_expected_reward}
\end{table}

\begin{figure}[t!]
    \centering
    \includegraphics[width=0.78\linewidth]{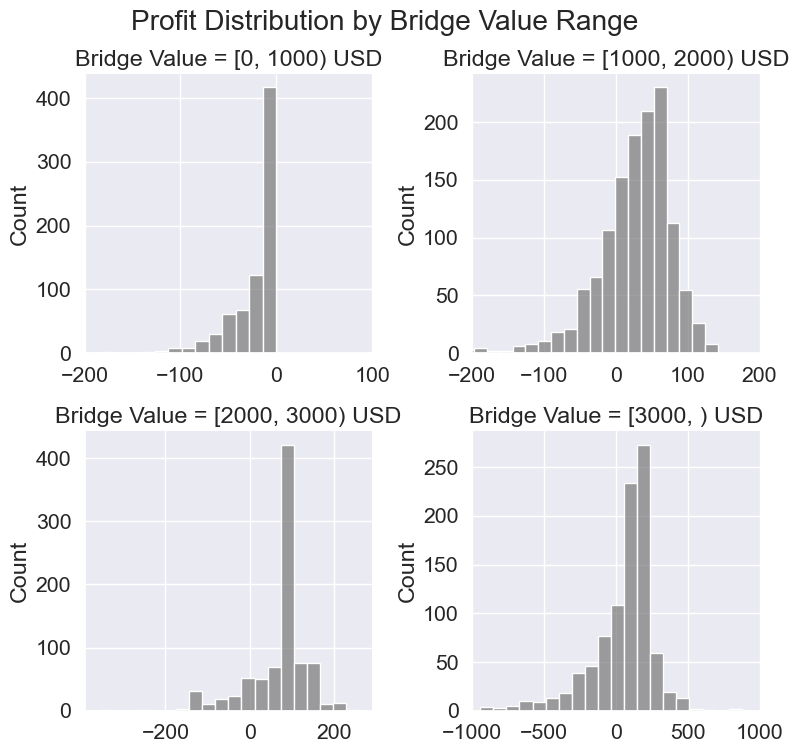}
    \caption{Address-level profit distribution by bridge value interval.}
    \label{fig:net_profit_dist_bridge_value}
\end{figure}

The results are presented in Figure \ref{fig:reward_fee_dist}, which displays the total reward and fee distribution across the various attacker groups ranked by profit. 
The first plot reveals that out of all the groups analyzed, 104 groups (69.3\%) exhibit a positive net profit, where the total reward exceeds the total fees. 
Conversely, the remaining groups demonstrate a negative profit, which indicates that the exploitation strategy was economically unfeasible for certain groups.
The majority of the groups' total profits fall within a relatively modest range, with most profits concentrated below \$10,000, while some notable outlier groups reach total rewards exceeding \$40,000.
The mean reward and fee per address demonstrate that on an individual level, the average per-address reward for most groups does not exceed \$200, and the highest expected return is below \$350 as demonstrated in Table \ref{tab:address_high_expected_reward}.
These findings provide an assessment of profit scale and economic feasibility at both the group and address levels.
In addition, Figure \ref{fig:net_profit_dist_bridge_value} illustrates the distribution of profits across various bridge value intervals for addresses. 
As the bridge value increases, the profit distribution becomes more dispersed, with higher bridge values showing a wider range of both positive and negative profits, demonstrating a greater variability in returns and costs for larger transactions.

\section{Airdrop Game theory Modeling}

The section discusses a sequential game theory model we proposed, which captures the intricate interactions among organizers, attackers, and bounty hunters (who report others to obtain rewards). 
By structuring the problem as a four-stage game, our model incorporates the strategic decision-making processes and private information of each participant. 
This framework considers uncertainties in the number of attackers and the private cost of bounty hunters. 
Utilizing contract theory, incentive-compatible contracts are designed to align the private incentives of bounty hunters with the goals of the organizer. 
We have a contracts setup to ensure that bounty hunters truthfully reveal their private information regarding detection capabilities and costs. 
By deriving optimal reward structures and task complexities, we minimize the organizer's expenses while maximizing detection efficacy. 
This contribution offers practical guidelines for organizers to structure contracts that motivate bounty hunters to participate.
    
We analytically derive the optimal self-report reward ratio (\( \pi_s^* \)) that organizers should offer to incentivize Sybil attackers to disclose their malicious activities. 
By balancing the benefits of reducing undetected attacks against the costs associated with rewarding self-reporting, we provide a formula for determining the optimal incentive. 
The result enables organizers to strategically set rewards that encourage attackers to self-report, reducing the number of undetected attacks and enhancing the overall fairness of the airdrop.

\subsection{Overview}
We consider a set of eligible addresses \( \mathcal{N} \triangleq \{1, 2, \ldots, N\} \) in the system, including both regular users and attackers. 
The total number of eligible addresses \( N \) is known, but only the expected number of attackers \( \mathbb{E}[N_h] \) is available.
The exact composition between regular users and attackers is uncertain. 
Each attacker incurs a public cost \( c_i \) for executing their attack. Since attackers are not involved in contract setting and their costs are publicly known, we do not classify them into types.
In contrast, bounty hunters are tasked with detecting attackers and possess two types of information: their detection capability \( \theta_j \) and cost structure \( C_j(\alpha_{b,j}) \). 
The total number of bounty hunters \( M \) is known, as they are required to register before the contract design for bounty reward phase. 
Each bounty hunter \( j \) is characterized by \( (\theta_j, C_j(\alpha_{b,j})) \) and is referred to as a type \( j \) bounty hunter. 
The set of \( M \) bounty hunters is partitioned into \( \mathcal{J} \triangleq \{1, 2, \ldots, J\} \) types, where each type \( j \in \mathcal{J} \) contains exactly one bounty hunter, satisfying \( J = M \). 
While the total number of bounty hunters and their distribution across types are public information, each bounty hunter’s specific type remains private.
The private nature of bounty hunters' information makes predicting their strategies challenging for the organizer. To address this, we propose a contract mechanism that elicits truthful information and aligns bounty hunters' incentives with the organizer's goal of detecting attackers.
Overall, the model consists of four distinct stages inspired by real-world scenarios \cite{layerzero_2024_addressing}:

\noindent \textbf{Stage 1 (Self-Report Reward Policy):}  
The organizer sets the self-report reward ratio \( \pi_s \in [0, 1] \), which determines the reward for attackers who choose to self-report their malicious activities.

\noindent \textbf{Stage 2 (Attackers' Self-Report Strategy):}  
Each attacker decides whether to self-report (\( x_i = 1 \)) or not (\( x_i = 0 \)), based on their cost \( c_i \) and the reward \( \pi_s \cdot R_i \) offered for self-reporting. The decision is influenced by the perceived detection probability \( \hat{p}_i \).

\noindent \textbf{Stage 3 (Bounty Reward Mechanism):}  
The organizer announces a menu of contracts \( \mathcal{C} = \{ (\alpha_{b,j}, r_{b,j}) \}_{j \in \mathcal{J}} \) to bounty hunters. Each contract specifies the complexity of detection tasks \( \alpha_{b,j} \) and the corresponding rewards \( r_{b,j} \) for successful detections.

\noindent \textbf{Stage 4 (Bounty Hunters' Strategy):}  
Bounty hunters select and sign contracts from the offered menu \( \mathcal{C} \) based on their private information \( (\theta_j, C_j(\alpha_{b,j})) \). Their objective is to maximize their individual utilities.

\subsection{Stage 1: Self-Report Reward Policy}
The organizer's utility \( U_o \) is the net benefit from mitigating undetected Sybil attacks, calculated as the value derived from reducing such attacks minus the total costs. Formally, $U_o = V_o - C_o(\pi_s)$ \label{U_o1}, where the value \( V_o \) is given by \( V_o = a \left(1 - \mathbb{E}[N_u^{(1)}] / N\right) T \) \label{V_o}, with \( a > 0 \) representing the value per reduction in undetected attacks, \( T \) the total tokens available for distribution, and \( N \) the total number of users (honest and attackers).
The total cost \( C_o(\pi_s) \) comprises rewards for self-reporting attackers and detection costs for non-self-reporting accounts:
\begin{equation}
    \label{C_o}
    C_o(\pi_s) = \pi_s \mathbb{E}[N_{\text{sr}}] \overline{R}_{\text{sr}} + \kappa \left(N - \mathbb{E}[N_{\text{sr}}]\right),
\end{equation}
where \( \pi_s \) is the self-report reward ratio, \( \overline{R}_{\text{sr}} = \frac{1}{\mathbb{E}[N_{\text{sr}}]} \sum_{i=1}^{\mathbb{E}[N_{\text{sr}}]} R_i \) is the average reward per self-reporting attacker, and \( \kappa > 0 \) denotes the detection cost per non-self-reporting account.
Post self-reporting decisions, the expected number of undetected attackers is
\begin{equation}
    \label{E_Nu}
    \mathbb{E}[N_u^{(1)}] = (1 - p_o) \mathbb{E}[N_h] (1 - \Sigma),
\end{equation}
where \( p_o \) is the probability of detecting an attacker, \( \mathbb{E}[N_h] \) is the expected number of attackers, and \( \Sigma = \frac{\mathbb{E}[N_{\text{sr}}]}{\mathbb{E}[N_h]} \) represents the self-reporting rate.

Substituting \(\mathbb{E}[N_u^{(1)}]\) from Equation \eqref{E_Nu} into the utility function, the organizer's utility in terms of \( \Sigma \) is

\begin{equation}
    \label{U_o_new}
    \resizebox{1.0\linewidth}{!}{$
    U_o = a \left(1 - \frac{(1 - p_o) \mathbb{E}[N_h] (1 - \Sigma)}{N}\right) T - \pi_s \mathbb{E}[N_{\text{sr}}] \overline{R}_{\text{sr}} - \kappa \left(N - \mathbb{E}[N_{\text{sr}}]\right)
    $}
\end{equation}

\noindent \textbf{Organizer Determines Self-Report Reward Ratio}:
The organizer determines the self-report reward ratio \( \pi_s \) to incentivize attackers to disclose their malicious activities. The organizer aims to maximize their utility \( U_o \), as expressed in Equation~\ref{U_o_new}. By solving the organizer's objective function in Stage 1, we derive the following result:
    
\begin{proposition}
\label{prop:optimal_pi_s}
The optimal self-report reward ratio \( \pi_s^* \) in Stage 1 is given by:
\[
\pi_s^* = \dfrac{1}{2} \left( 1 - p_o - \dfrac{a (1 - p_o) T - \kappa N}{\overline{R}_{\text{sr}} N} \right)
\]
\end{proposition}
    
Proof of Proposition~\ref{prop:optimal_pi_s} is at in Appendix~\ref{appendix:stage_1}.
The optimal self-report reward ratio \( \pi_s^* \) balances attackers' incentives to self-report with the organizer's costs. By setting \( \pi_s^* \) appropriately, the organizer encourages disclosure without excessive expenses. If the detection cost \( \kappa \) is high, an increase in \( \pi_s^* \) ensures adequate incentives for reporting. As the number of attackers \( N \) grows, the organizer can adjust \( \pi_s^* \) to maintain scalability and optimal incentives.

\subsection{Stage 2: Attackers' Self-Report Strategy}

Each Sybil attacker \( i \) seeks to maximize their expected utility \( U_i \) by deciding whether to self-report: $U_i = x_i (\pi_s R_i) + (1 - x_i) \left( (1 - \hat{p}_i) R_i \right) - c_i$, \label{U_i}
where \( x_i \in \{0, 1\} \) indicates the decision to self-report (\( x_i = 1 \)) or not (\( x_i = 0 \)). If \( x_i = 1 \), the attacker receives the self-report reward \( \pi_s R_i \) and incurs the cost \( c_i \). If \( x_i = 0 \), the attacker receives the expected reward \( (1 - \hat{p}_i) R_i \) and still incurs the cost \( c_i \).

The perceived detection probability for Sybil attacker \( i \), denoted by \( \hat{p}_i \), is modeled as: $\hat{p}_i(\Sigma) = p_o + \phi \Sigma$, \label{hat_p_i}, where \( p_o \) is the baseline detection probability, \( \phi > 0 \) captures the influence of other attackers' decisions on the detection probability. \( \Sigma \) represents the fraction of attackers who choose to self-report, considering the externalities among attackers\cite{davis1962externalities}. Specifically, $\Sigma = \frac{1}{\mathbb{E}[N_h]} \sum_{j=1}^{N} x_j$, where  \( x_j \) is the self-reporting decision of attacker \( j \).



In Stage 2, attackers make strategic decisions regarding whether to self-report their activities. The set of players consists of all attackers, denoted by \( \mathcal{S} = \{1, 2, \dots, N\} \), where \( N \) is the total number of attackers. Each attacker \( i \in \mathcal{S} \) faces a binary decision: either to self-report their malicious behavior, represented by \( x_i = 1 \), or to not self-report, represented by \( x_i = 0 \).

\noindent \textbf{Payoff Function}: Each Sybil attacker \(i\) aims to maximize their expected utility \(U_i(x_i, \mathbf{x}_{-i})\), where \(\Sigma\) is the fraction of attackers who choose to self-report:
\begin{equation}
    \label{U_i_stage2}
    U_i(x_i, \mathbf{x}_{-i}) = x_i \left( \pi_s R_i - c_i \right) + (1 - x_i) \left( (1 - \hat{p}_i(\Sigma)) R_i - c_i \right)
\end{equation}

\subsubsection{Optimal Decision for Each Attacker:}
Each attacker \(i\) determines their optimal decision by comparing the utilities of self-reporting and not self-reporting. Attacker \(i\) will choose \(x_i = 1\) (self-report) if: $U_i(1, \mathbf{x}_{-i}) \geq U_i(0, \mathbf{x}_{-i})$.
The attacker decision rule simplifying is provided in Appendix \ref{appendix:stage_2}. As a result,  we have $\hat{p}_i(\Sigma) \geq 1 - \pi_s$.
Since \(\hat{p}_i(\Sigma)\) increases with \(\Sigma\), as more attackers self-report (\(\Sigma\) increases), the condition \(\hat{p}_i(\Sigma) \geq 1 - \pi_s\) is more likely to be satisfied, incentivizing attacker \(i\) to self-report.  Attackers assess their expected gains from self-reporting against the costs.

\subsection{Stage 3: Bounty Reward Mechanism}

The organizer designs a menu of contracts  \( \mathcal{C} = \{ (\alpha_{b,j}, r_{b,j}) \}_{j \in \mathcal{J}} \) to incentivize bounty hunters to exert appropriate effort in detecting Sybil attackers while minimizing costs. The organizer's utility in this stage depends on the expected number of undetected Sybil attackers after bounty hunters perform their detection tasks and the total rewards allocated to them.

The expected number of undetected Sybil attackers after bounty hunters have executed their detection tasks is given by
\(
    \label{E_Nu3}
    \mathbb{E}[N_u^{(3)}] = \mathbb{E}[N_u^{(1)}] -  p_{b,j} \sum_{j \in \mathcal{J}} |\mathcal{D}_j|,
\)
where \( \mathbb{E}[N_u^{(1)}] \) is the expected number of undetected Sybil attackers after self-reporting and the organizer's initial detection, and \( |\mathcal{D}_j| = \frac{\mathbb{E}[N_u^{(1)}]}{M} \) represents the expected number of detections by bounty hunter \( j \). Here, \( M \) is the total number of bounty hunters, and \( p_{b,j} \) is the detection probability for bounty hunter \( j \).

The detection probability \( p_{b,j} \) for bounty hunter \( j \) is modeled using a sigmoid function to ensure it lies within the interval (0, 1):
\(
    p_{b,j} = \sigma(\theta_j - \gamma \alpha_{b,j}),
\)
where \( \sigma(z) = \frac{1}{1 + e^{-z}} \) is the sigmoid function, \( \theta_j \) denotes the detection capability of bounty hunter \( j \), and \( \gamma > 0 \) captures the inverse relationship between task complexity \( \alpha_{b,j} \) and detection probability.

The total cost incurred by the organizer for the contracts is the sum of the rewards allocated to all bounty hunters:
\(
C_o^{(3)} = \sum_{j \in \mathcal{J}} r_{b,j},
\)
where \( r_{b,j} \) is the reward allocated to bounty hunter \( j \). This cost function represents the organizer's expenditure on incentivizing bounty hunters to effectively perform their detection tasks. Finally, the organizer's utility in this stage depends on the expected number of undetected Sybil attackers after bounty hunters perform their detection tasks and the total rewards allocated to them.
Here, we show the optimal reward policy for organizers.

\begin{lemma}[Optimal Rewards for Given Task Complexities]
\label{lemma2}
For any given set of task complexities \( \boldsymbol{\alpha}_b = \{ \alpha_{b,j} \}_{j \in \mathcal{J}} \), the unique optimal reward \( r_{b,j}^* \) for a bounty hunter of type \( j \) is given by:
\[
\resizebox{0.48\textwidth}{!}{$
r_{b,j}^* =
\begin{cases}
C_j(\alpha_{b,j}), & \text{if } j = J, \\
C_j(\alpha_{b,j}) + \sum\limits_{m=j+1}^J \left[ C_m(\alpha_{b,m}) - C_{m-1}(\alpha_{b,m}) \right], & \text{if } j = 1, \dots, J-1.
\end{cases}
$}
\]
\end{lemma}

\noindent The proof of Lemma~\ref{lemma2} is detailed in Appendix~\ref{appendix:stage_3}. By calculating the optimal rewards, the organizer minimizes expenses while maintaining effective detection rates. Optimal rewards ensure that bounty hunters are adequately compensated for their efforts, aligning their interests with the organizer's goals. Lemma~\ref{lemma2} shows that all bounty hunter types except type \( J \) receive positive expected payoffs, while type \( J \) hunters earn nothing. This difference represents the information rent resulting from information asymmetry \cite{bolton2004contract}.

\begin{proposition}[Optimal Reward Ratio \( \pi_{b,j}^* \)]
\label{prop1}
Given the optimal rewards \( \{ r_{b,j}^* \}_{j \in \mathcal{J}} \) derived in Lemma~\ref{lemma2}, the unique optimal reward ratio \( \pi_{b,j}^* \) for a bounty hunter of type \( j \) is:
\[
\pi_{b,j}^* = r_{b,j}^* \cdot \left(1 + e^{-(\theta_j - \gamma \alpha_{b,j})}\right) \cdot \left(\frac{1}{\sum_{i \in \mathcal{D}_j} R_i}\right),~\forall j \in \mathcal{J}.
\]
\end{proposition}

\noindent See the proof of Proposition~\ref{prop1} in Appendix~\ref{appendix:stage_3}.
Besides, we also present a numerical algorithm to derive the optimal task complexities \( \alpha_{b,j}^* \) to minimize the objective function, with the proof of the following theorem:

\begin{theorem}[Contract Feasibility]
The contract, denoted as: \( \mathcal{C^*} = \{ (\alpha_{b,j}^*, r_{b,j}^*) \}_{j \in \mathcal{J}} \) is feasible, i.e., satisfies both Incentive Compatibility (IC) and Individual Rationality (IR) constraints defined in Definition \ref{def: ic} and \ref{def: ir} in Appendix \ref{appendix:stage_3}. 
\end{theorem}

\noindent Overall, the results provide practical guidelines for organizers to align incentives across all parties involved.

\subsection{Stage 4: Bounty Hunters' Strategy}

In this stage, we model the utility \( U_{b,j} \) for bounty hunter \( j \) upon selecting contract \( (\alpha_{b,j}, r_{b,j}) \). The utility is defined as:
\(
    U_{b,j} = r_{b,j} - C_j(\alpha_{b,j}).
\)
The reward rate \( r_{b,j} \) for bounty hunter \( j \) is determined by
\(
    r_{b,j} = \pi_{b,j} \cdot p_{b,j} \cdot \sum_{i \in \mathcal{D}_j} R_i,
\)
where \( \pi_{b,j} \) is the reward ratio to be determined, \( p_{b,j} \) is the detection probability for bounty hunter \( j \) as defined in Stage $3$, and \( \sum_{i \in \mathcal{D}_j} R_i \) represents the total rewards from detected Sybil attackers by bounty hunter \( j \).
The cost incurred by bounty hunter \( j \) for handling detection tasks is modeled as:
\(
    C_j(\alpha_{b,j}) = c_j \cdot \alpha_{b,j} + d_j,
\)
where \( c_j \) is the variable cost coefficient associated with the task complexity \( \alpha_{b,j} \), and \( d_j \) is the fixed cost incurred by bounty hunter \( j \) regardless of the task complexity.

Bounty hunters select and sign contracts from the menu offered by the organizer based on their private information \( (\theta_j, C_j(\alpha_{b,j})) \). Their objective is to maximize individual utilities while adhering to their cost structures, i.e., each bounty hunter \( j \) selects the contract that maximizes their utility:
\(
\alpha_{b,j}^*, r_{b,j}^* = \argmax U_{b,j},
\)
subject to the contract menu designed by the organizer and the incentive compatibility and participation constraints outlined in Stage $3$.

\section{Conclusion}
This study analyzed airdrop hunter behaviors on reported hunter groups of Hop Protocol and LayerZero, identifying key patterns like funding, sequential bridging, and uniformity. 
Our profit modeling evaluated the expected rewards and costs, revealing the distribution of group level positive or negative net profits.
We also proposed a game-theory model to optimize reward policies, balancing detection incentives with minimizing organizer costs. These insights offer frameworks to enhance the resilience of future airdrop distributions.

\clearpage

\bibliographystyle{ieeetran}
\bibliography{reference.bib}

\titleformat{\section}[block]
  {\normalfont\Large\bfseries}
  {Appendix \thesection:}{1em}{}
  
\appendix

\section{Pattern Measure Algorithms}
\label{pattern_measure_alg}

\vspace{-1em}

\begin{algorithm}[!htbp]
\caption{Initial Funder and Common Receiver Detection}
\begin{algorithmic}[1]
\State \textbf{Input:} $G = \{A_1, A_2, \dots, A_n\}$: A group of attacker addresses, $T_{tx}$: Bridge transactions
\State \textbf{Output:} $\mathcal{F}$, $\mathcal{R}$
\State \textbf{Procedure:}
\State \hspace{1em} $T_{to} \leftarrow \{T \in T_{tx} \mid T_{\text{to}} \in G\}$ 
\State \hspace{1em} $T_{from} \leftarrow \{T \in T_{tx} \mid T_{\text{from}} \in G\}$ 
\State \hspace{1em} $F_{\text{unique}} \leftarrow \text{Group by } T_{\text{from}} \text{ and count unique } T_{\text{to}}$ 
\State \hspace{1em} $R_{\text{unique}} \leftarrow \text{Group by } T_{\text{to}} \text{ and count unique } T_{\text{from}}$ 
\State \hspace{1em} $A_{\text{funder}} \leftarrow \arg\max(F_{\text{unique}})$
\State \hspace{1em} $p_{\text{funded}} \leftarrow \frac{F_{\text{unique}}[A_{\text{funder}}]}{|G|} $ \Comment{Percent of addresses funded}
\State \hspace{1em} $A_{\text{receiver}} \leftarrow \arg\max(R_{\text{unique}})$ 
\State \hspace{1em} $p_{\text{received}} \leftarrow \frac{R_{\text{unique}}[A_{\text{receiver}}]}{|G|} $ \Comment{Percent of addresses sent}
\State \hspace{1em} $\mathcal{F} \leftarrow (A_{\text{funder}}, p_{\text{funded}})$,  $\mathcal{R} \leftarrow (A_{\text{receiver}}, p_{\text{received}})$
\State \textbf{Return} $\mathcal{F}, \mathcal{R}$

\end{algorithmic}
\label{alg:initial_funder_common_receiver_detection}
\end{algorithm}

\vspace{-0.6em}

\begin{algorithm}[!htbp]
\caption{Sequential Transfer Detection}
\begin{algorithmic}[1]

\State \textbf{Input:} $G = \{A_1, A_2, \dots, A_n\}$: A group of attacker addresses, $\varepsilon$: Value threshold, $\Delta_t$: Time threshold
\State $T_{tx}$: All bridge transactions

\State \textbf{Output:} $\mathcal{S}$: Sequential transfer chains

\State \textbf{Procedure:}

\State \textbf{1:} \text{For each } $A_i \in G$: 
\State \hspace{1em} $T_{tx}^{A_i} \leftarrow \{T \in T_{tx} \mid A_i \text{ is in } T\}$ \Comment{Filter transactions by address}
\State \hspace{1em} $V_{\text{out}}^{A_i}, V_{\text{in}}^{A_i} \leftarrow \text{Extract}(T_{tx}^{A_i})$ \Comment{Get outgoing and incoming values}

\State \textbf{2:} \text{For each pair } $(A_{\text{from}}, A_{\text{to}})$:
\State \hspace{1em} $\Delta t = |t_{\text{out}} - t_{\text{in}}| \leq \Delta_t$ \Comment{Check time difference}
\State \hspace{1em} $\Delta V = \left|\frac{V_{\text{out}} - V_{\text{in}}}{V_{\text{out}}}\right| \leq \varepsilon$ \Comment{Check value difference}
\State \hspace{1em} \text{Add } $(T_{tx}^{\text{out}}, T_{tx}^{\text{in}})$ if conditions hold \Comment{Store valid transactions}

\State \textbf{3:} $\mathcal{C} \leftarrow \{\}$ \Comment{Initialize chain set}

\State \text{For each } $T_{tx}^{\text{out}} \in T_{tx}$:
\State \hspace{1em} $T_{tx}^{\text{in}} \leftarrow \{T_{tx} \mid T_{tx}^{\text{in}} \text{ satisfies Step 2}\}$ \Comment{Find next valid transaction}
\State \hspace{1em} $C \leftarrow C \cup T_{tx}^{\text{in}}$ \Comment{Append to chain}
\State \hspace{1em} \text{Repeat until no valid $T_{tx}$}
\State \hspace{1em} $\mathcal{C} \leftarrow \mathcal{C} \cup C$ \Comment{Add chain to set}

\State \textbf{4:} For each $C \in \mathcal{C}$:
\State \hspace{1em} $(A_{\text{start}}, A_{\text{end}}) \leftarrow \text{start and end of } C$
\State \hspace{1em} $(V_{\text{start}}, V_{\text{end}}) \leftarrow \text{values of } C$
\State \hspace{1em} $(T_{\text{start}}, T_{\text{end}}) \leftarrow \text{timestamps of } C$

\State \textbf{Return} $\mathcal{S} \leftarrow \mathcal{C}$ \Comment{Return all sequential transfers (chains)}

\end{algorithmic}
\label{alg:sequential_transfer_detection}
\end{algorithm}

\begin{algorithm}[htbp!]
\caption{Uniformity Calculation }
\begin{algorithmic}[1]

\State \textbf{Input:} $G = \{A_1, A_2, \dots, A_n\}$: A group of attacker addresses
\State $\varepsilon_T, \varepsilon_F$: Thresholds for count and volume uniformity, $T_{tx}$: All bridge transactions

\State \textbf{Output:} $\mathcal{U}$: Uniformity scores for each group $(U_T, U_F)$

\State \textbf{Procedure:}

\State \textbf{1:} \text{For each } $A_i \in G$: 
\State \hspace{1em} $C_{\text{count}}(A_i) \leftarrow 0, V_{\text{volume}}(A_i) \leftarrow 0$ 
\State \hspace{1em} \text{For each } $T \in T_{tx}$:
\State \hspace{2em} $C_{\text{count}}(A_i) \leftarrow C_{\text{count}}(A_i) + 1$
\State \hspace{2em} $V_{\text{volume}}(A_i) \leftarrow V_{\text{volume}}(A_i) + V$ 

\State \textbf{2:} \text{For each group } $A_i$:
\State \hspace{1em} \text{Reshape } $C_{\text{count}}, V_{\text{volume}}$ \text{ to vectors } $X_T, X_F$ 
\State \hspace{1em} $n_T \leftarrow \min(3, |X_T|), n_F \leftarrow \min(3, |X_F|)$ 
\State \hspace{1em} \text{Fit means to } $X_T, X_F$

\State \textbf{3:} \text{Calculate uniformity:}
\State \hspace{1em} \text{For counts } $X_T$: 
\State \hspace{2em} $U_T \leftarrow \frac{|\{x_T \in X_T \mid |x_T - \mu_T| < \varepsilon_T\}|}{|X_T|}$ 
\State \hspace{1em} \text{For volumes } $X_F$: 
\State \hspace{2em} $U_F \leftarrow \frac{|\{x_F \in X_F \mid |x_F - \mu_F| < \varepsilon_F\}|}{|X_F|}$ 

\State \textbf{4:} $\mathcal{U} \leftarrow \mathcal{U} \cup (U_T, U_F)$ 
\Comment{Append uniformity tuple for each group}

\State \textbf{5:} \textbf{Return} $\mathcal{U}$ for each $A_i$

\end{algorithmic}
\label{alg:uniformity_detection}
\end{algorithm}

\section{Game Theory Model Detail and Proofs}
\subsection{Stage 1: Self-Report Reward Policy}
\label{appendix:stage_1}
\

\vspace{0.6em}
\noindent \textit{\textbf{Proof of Proposition~\ref{prop:optimal_pi_s}}}

\label{appendix:proofs_pro3}

\begin{proof}
To determine the optimal self-report reward ratio \( \pi_s^* \), we analyze the organizer's utility function \( U_o(\pi_s) \) and optimize it with respect to \( \pi_s \).

Starting with the equilibrium condition where attackers are indifferent between self-reporting and not self-reporting:
\[
\Sigma = \dfrac{1 - \pi_s - p_o}{\phi}
\]
where \( \Sigma \) represents the proportion of attackers who self-report.

\noindent Substituting \( \Sigma = \dfrac{1 - \pi_s - p_o}{\phi} \) into the organizer's utility function:
{\small
\[
U_o(\pi_s) = a \left( 1 - \dfrac{(1 - p_o) \mathbb{E}[N_h] \left( 1 - \dfrac{1 - \pi_s - p_o}{\phi} \right)}{N} \right) T 
\]
\[
- \pi_s \mathbb{E}[N_{\text{sr}}] \overline{R}_{\text{sr}} - \kappa \left( N - \mathbb{E}[N_{\text{sr}}] \right)
\]
}
Given that \( \mathbb{E}[N_{\text{sr}}] = \dfrac{\mathbb{E}[N_h] (1 - \pi_s - p_o)}{\phi} \), we substitute back:

{\scriptsize
\[
U_o(\pi_s) = a \left(1 - \dfrac{(1 - p_o)(1 - \pi_s - p_o)}{\phi N} \mathbb{E}[N_h] \right) T 
\]
\[
- \dfrac{\mathbb{E}[N_h] \overline{R}_{\text{sr}}}{\phi} \pi_s (1 - \pi_s - p_o) - 
\kappa \left( N - \dfrac{\mathbb{E}[N_h] (1 - \pi_s - p_o)}{\phi} \right)
\]
}
Simplifying the expression:
{\scriptsize
\[
U_o(\pi_s) = a \left(1 - \dfrac{(1 - p_o)(1 - \pi_s - p_o)}{\phi N} \mathbb{E}[N_h] \right) T 
\]
\[- \dfrac{\mathbb{E}[N_h] \overline{R}_{\text{sr}}}{\phi} \pi_s (1 - \pi_s - p_o) - \kappa \left(N - \dfrac{\mathbb{E}[N_h] (1 - \pi_s - p_o)}{\phi}\right)
\]
}

To find the optimal \( \pi_s^* \), we take the derivative of \( U_o(\pi_s) \) with respect to \( \pi_s \) and set it to zero:
\[
\dfrac{dU_o}{d\pi_s} = 0
\]

Calculating the derivative:
{\scriptsize
\begin{align*}
\dfrac{dU_o}{d\pi_s} &= a \left( -\dfrac{(1 - p_o)}{\phi N} \mathbb{E}[N_h] \right) T - \dfrac{\mathbb{E}[N_h] \overline{R}_{\text{sr}}}{\phi} \left( (1 - \pi_s - p_o) - \pi_s \right) + \dfrac{\kappa \mathbb{E}[N_h]}{\phi} \\
&= -\dfrac{a (1 - p_o) T \mathbb{E}[N_h]}{\phi N} - \dfrac{\mathbb{E}[N_h] \overline{R}_{\text{sr}}}{\phi} (1 - 2\pi_s - p_o) + \dfrac{\kappa \mathbb{E}[N_h]}{\phi} \\
&= 0
\end{align*}
}

Multiplying both sides by \( \phi \) and dividing by \( \mathbb{E}[N_h] \) (assuming \( \mathbb{E}[N_h] \neq 0 \)):
\[
- a \dfrac{(1 - p_o) T}{N} - \overline{R}_{\text{sr}} (1 - 2\pi_s - p_o) + \kappa = 0
\]
\[
\overline{R}_{\text{sr}} (1 - 2\pi_s - p_o) = -a \dfrac{(1 - p_o) T}{N} + \kappa
\]
\[
1 - 2\pi_s - p_o = \dfrac{-a (1 - p_o) T + \kappa N}{\overline{R}_{\text{sr}} N}
\]
\[
-2\pi_s = \dfrac{-a (1 - p_o) T + \kappa N}{\overline{R}_{\text{sr}} N} - 1 + p_o
\]
\[
2\pi_s = 1 - p_o - \dfrac{a (1 - p_o) T - \kappa N}{\overline{R}_{\text{sr}} N}
\]
\[
\pi_s^* = \dfrac{1}{2} \left(1 - p_o - \dfrac{a (1 - p_o) T - \kappa N}{\overline{R}_{\text{sr}} N}\right)
\]
This expression yields the optimal self-report reward ratio \( \pi_s^* \) that balances the incentives for Sybil attackers to self-report against the associated costs and detection probabilities.
\end{proof}

\subsection{Stage 2: Attackers' Self-Reporting Strategy}
\label{appendix:stage_2}

\begin{proposition}
\label{prop:supermodular_game}
The subgame in Stage 2, where Sybil attackers decide whether to self-report their malicious activities, is a supermodular game. Consequently, there exists at least one pure strategy Nash Equilibrium.
\end{proposition}

Understanding that the subgame is supermodular ensures the existence of a stable equilibrium where attackers' strategies complement each other. This allows organizers to anticipate and influence attacker behavior. Attacker decision rule simplifying:
\begin{align*}
& \pi_s R_i - c_i \geq (1 - \hat{p}_i(\Sigma)) R_i - c_i \\
\Rightarrow & \pi_s R_i \geq (1 - \hat{p}_i(\Sigma)) R_i \\
\Rightarrow & \pi_s \geq 1 - \hat{p}_i(\Sigma) \\
\Rightarrow & \hat{p}_i(\Sigma) \geq 1 - \pi_s
\end{align*}

\noindent \textit{\textbf{Proof of Proposition~\ref{prop:supermodular_game}}}
\label{appendix:proofs_pro2}
\begin{proof}
To demonstrate that the subgame is a supermodular game, we verify the necessary conditions for supermodularity.

Firstly, each player's strategy set is \( \{0, 1\} \), which forms a complete lattice. This satisfies the requirement that strategy sets are lattices.

Secondly, we examine the payoff function \( U_i(x_i, \mathbf{x}_{-i}) \) for increasing differences in \( (x_i, \mathbf{x}_{-i}) \). Consider two strategy profiles \( \mathbf{x}_{-i}' \geq \mathbf{x}_{-i} \), implying that more attackers choose to self-report in \( \mathbf{x}_{-i}' \) than in \( \mathbf{x}_{-i} \). Define \( \Sigma = \frac{1}{N}(x_i + S_{-i}) \) and \( \Sigma' = \frac{1}{N}(x_i + S_{-i}') \), where \( S_{-i} = \sum_{j \neq i} x_j \) and \( S_{-i}' = \sum_{j \neq i} x_j' \), with \( S_{-i}' \geq S_{-i} \).

The utility difference when player \( i \) switches from not reporting to reporting under the two profiles is:

\[
\Delta U_i = \left[ U_i(1, \mathbf{x}_{-i}') - U_i(0, \mathbf{x}_{-i}') \right] - \left[ U_i(1, \mathbf{x}_{-i}) - U_i(0, \mathbf{x}_{-i}) \right]
\]

Substituting the payoff functions:

\begin{align*}
\Delta U_i &= \left[ (\pi_s R_i - c_i) - \left( (1 - \hat{p}_i(\Sigma')) R_i - c_i \right) \right] \\ 
& \quad\quad\quad - \left[ (\pi_s R_i - c_i) - \left( (1 - \hat{p}_i(\Sigma)) R_i - c_i \right) \right] \\
&= \left[ \pi_s R_i - (1 - \hat{p}_i(\Sigma')) R_i \right] - \left[ \pi_s R_i - (1 - \hat{p}_i(\Sigma)) R_i \right] \\
&= \left[ \hat{p}_i(\Sigma') - \hat{p}_i(\Sigma) \right] R_i
\end{align*}

Given that \( \Sigma' \geq \Sigma \) and assuming \( \hat{p}_i(\Sigma) \) is an increasing function of \( \Sigma \), it follows that \( \hat{p}_i(\Sigma') \geq \hat{p}_i(\Sigma) \). Therefore, \( \Delta U_i \geq 0 \), which confirms that the payoff function exhibits increasing differences in \( (x_i, \mathbf{x}_{-i}) \).

Since both conditions for supermodularity are satisfied—the strategy sets form complete lattices and the payoff functions possess increasing differences—the game is classified as a supermodular game.

By \cite{supermodular_game_ref}, every finite supermodular game possesses at least one pure strategy Nash Equilibrium. Therefore, the subgame in Stage 2 has at least one pure NE.
\end{proof}

\subsection{Stage 3: Bounty Reward Mechanism}
\label{appendix:stage_3}
\begin{definition}[Incentive Compatibility (IC)]
\label{def: ic}
A contract mechanism satisfies \textbf{IC} if each bounty hunter maximizes their utility by truthfully selecting the contract designed for their own type, rather than deviating to a contract intended for another type. Formally, for every pair of contract types \( j, k \in \mathcal{J} \) with \( k \neq j \), the following condition must hold:
\[
 r_{b,j} - C_j(\alpha_{b,j}) \geq r_{b,k} - C_j(\alpha_{b,k})
\]
\end{definition}

\begin{definition}[Individual Rationality (IR)]
\label{def: ir}
A contract mechanism satisfies \textbf{IR} if every bounty hunter derives non-negative utility from participating in the contract. Formally, for every \( j \in \mathcal{J} \), the following condition must be satisfied:
\[
r_{b,j} - C_j(\alpha_{b,j}) \geq 0
\]
\end{definition}

The organizer seeks to minimize the expected number of undetected Sybil attackers \( \mathbb{E}[N_u^{(3)}] \) relative to the total rewards allocated to bounty hunters. Formally, the organizer's objective is:
\[
\min_{\mathcal{C}} \quad \frac{\mathbb{E}[N_u^{(3)}]}{\sum_{j \in \mathcal{J}} r_{b,j} |\mathcal{D}_j|}
\]
Subject to:
\[
 r_{b,j} - C_j(\alpha_{b,j}) \geq r_{b,k} - C_j(\alpha_{b,k}), \quad \forall j, k \in \mathcal{J}, j \neq k
\]
\[
r_{b,j} - C_j(\alpha_{b,j}) \geq 0, \quad \forall j \in \mathcal{J}
\]

\noindent \textbf{Optimal Reward and Reward Ratio}
To solve the above optimization problem, we first determine the optimal rewards. The feasibility of the contract, satisfying both IR and IC constraints, is established in Lemma~\ref{lemma1}.

\begin{lemma}[Feasibility of the Contract]
\label{lemma1}
A contract \( \mathcal{C} = \{ (\alpha_{b,j}, r_{b,j}) \}_{j \in \mathcal{J}} \) is feasible, meaning it satisfies both Individual Rationality (IR) and Incentive Compatibility (IC) constraints, if and only if the following conditions are satisfied:
\begin{enumerate}[label=(\alph*)]
    \item 
    \[
    U_{b,J} = r_{b,J} - C_J(\alpha_{b,J}) \geq 0
    \]
    
    \item
    \[
    r_{b,1} \geq r_{b,2} \geq \dots \geq r_{b,J} \geq 0
    \]
    \[
    \alpha_{b,1} \geq \alpha_{b,2} \geq \dots \geq \alpha_{b,J}
    \]
    
    \item 
    \[
    r_{b,j+1} - C_j(\alpha_{b,j+1}) 
    \leq 
    r_{b,j} - C_j(\alpha_{b,j})
    \]
    \[
    \leq
    r_{b,j+1} - C_{j+1}(\alpha_{b,j+1}),
 \forall j \in \{1, \dots, J-1\}
    \]
\end{enumerate}
\end{lemma}

The feasibility conditions ensure that the contracts are both attractive and truthful, preventing bounty hunters from misrepresenting their types. By adhering to these conditions, organizers can allocate rewards optimally, minimizing costs while maximizing detection efficacy. As for bounty hunters, the IC and IR constraints guarantee they are motivated to participate truthfully and exert appropriate effort.

\noindent \textbf{Optimal Task Complexities}
Following Lemma~\ref{lemma2}, the organizer determines the optimal task complexities \( \alpha_{b,j}^* \) to minimize the objective function:
\[
\min_{\{ \alpha_{b,j} \}_{j \in \mathcal{J}}} \quad \frac{\mathbb{E}[N_u^{(3)}]}{\sum_{j \in \mathcal{J}} r_{b,j}^*}
\]
Subject to:
\[
\alpha_{b,1} \geq \alpha_{b,2} \geq \dots \geq \alpha_{b,J} \geq 0
\]
The detailed derivation and optimization approach are elaborated below.
We present Algorithm~\ref{alg:optimal_alpha}, the organizer can numerically determine the optimal task complexities \( \alpha_{b,j}^* \) that minimize the objective function \( \frac{\mathbb{E}[N_u^{(3)}]}{\sum_{j} r_{b,j}^*} \) while satisfying the task complexity ordering constraint.
The time complexity of the algorithm is \(O \left( J \times \log \left( \frac{1}{\varepsilon} \right) \right)\), where: \(\varepsilon\) is the desired precision for \(\lambda\) in the root-finding method.
From the result, we conclude that the he optimal task complexities \( \alpha_{b,j}^* \) are designed to strike a balance between maximizing detection probabilities and minimizing costs. The algorithm ensures that this trade-off is optimized to achieve the lowest possible ratio of undetected attackers to total rewards.
Meanwhile, the ordering constrain ensures incentive compatibility and contract feasibility, preventing bounty hunters from misreporting their types and ensuring truthful participation.

\begin{algorithm}[t!]
\caption{Algorithm to Find Optimal \( \alpha_{b,j}^* \)}
\label{alg:optimal_alpha}
\begin{algorithmic}[1]
\Require Parameters \( \theta_j \), \( c_j \), \( M \), \( J \), \( \gamma \)
\Ensure Optimal task complexities \( \{ \alpha_{b,j}^* \} \)

\State Initialize \( \lambda \) with an initial guess
\Repeat
    \For{each type \( j = 1, 2, \dots, J \)}
        \State Compute \( a_j = \frac{ \lambda M (J - j + 1) c_j }{ \gamma } \)
        \If{ \( a_j > \frac{1}{4} \) }
            \State Adjust \( \lambda \) and \textbf{restart}
        \EndIf
        \State Compute \( p_{b,j} = \frac{1}{2} \left( 1 - \sqrt{1 - 4 a_j} \right) \)
        \State Compute \( \alpha_{b,j} = \frac{1}{\gamma} \left( \theta_j + \ln \left( \frac{1 - p_{b,j}}{p_{b,j}} \right) \right) \)
    \EndFor
    \State \textbf{Check} the ordering constraint:
    \State \( \alpha_{b,1} \geq \alpha_{b,2} \geq \dots \geq \alpha_{b,J} \)
    \If{ordering is satisfied}
        \State \textbf{Terminate}
    \Else
        \State Adjust \( \lambda \) (e.g., using a root-finding method)
    \EndIf
\Until{Ordering constraint is satisfied}
\State \Return \( \{ \alpha_{b,j}^* \} \)
\end{algorithmic}
\end{algorithm}

\vspace{0.7em}
\noindent \textit{\textbf{Proof of Lemma~\ref{lemma1}:  Feasibility of the Contract}}

\label{appendix:proofs_lemma1}

\begin{proof}
To demonstrate that a contract \( \mathcal{C} = \{ (\alpha_{b,j}, r_{b,j}) \}_{j \in \mathcal{J}} \) is feasible, meaning it satisfies both Individual Rationality (IR) and Incentive Compatibility (IC) constraints, if and only if conditions (a), (b), and (c) are satisfied.

\subsubsection{Necessary Conditions}

Assume that the contract \( \mathcal{C} \) satisfies the IR and IC constraints. We will show that conditions (a), (b), and (c) must hold.

\paragraph{Condition (a): \( U_{b,J} = r_{b,J} - C_J(\alpha_{b,J}) \geq 0 \)}

The IR constraint for the lowest type \( J \) requires:
\[
U_{b,J} = r_{b,J} - C_J(\alpha_{b,J}) \geq 0
\]
This directly corresponds to condition (a).

\paragraph{Condition (b): \( r_{b,1} \geq r_{b,2} \geq \dots \geq r_{b,J} \geq 0 \) and \( \alpha_{b,1} \\
\geq \alpha_{b,2} \geq 
\dots \geq \alpha_{b,J} \)}

To satisfy the IC constraints, the rewards and task complexities must be non-increasing with respect to the types. Assume types are ordered such that \( \theta_1 > \theta_2 > \dots > \theta_J \).

Suppose, for contradiction, that \( r_{b,j} < r_{b,j+1} \) for some \( j \). Then, type \( j \) could obtain higher utility by misreporting as type \( j+1 \):
\[
U_{b,j}(\alpha_{b,j+1}, r_{b,j+1}) = r_{b,j+1} - C_j(\alpha_{b,j+1}) > r_{b,j} - C_j(\alpha_{b,j})
\]
This violates the IC constraint. Therefore, rewards must satisfy:
\[
r_{b,1} \geq r_{b,2} \geq \dots \geq r_{b,J} \geq 0
\]

Similarly, if \( \alpha_{b,j} < \alpha_{b,j+1} \), a higher-type bounty hunter may prefer the contract intended for a lower type, violating IC. Thus, task complexities must satisfy:
\[
\alpha_{b,1} \geq \alpha_{b,2} \geq \dots \geq \alpha_{b,J}
\]

\paragraph{Condition (c): \( r_{b,j+1} - C_j(\alpha_{b,j+1}) \leq r_{b,j} - C_j(\alpha_{b,j}) \leq r_{b,j+1} \\
- C_{j+1}(\alpha_{b,j+1}) \)}

To prevent type \( j \) from misreporting as type \( j+1 \), we require:
\[
r_{b,j} - C_j(\alpha_{b,j}) \geq r_{b,j+1} - C_j(\alpha_{b,j+1})
\]
which implies:
\[
r_{b,j} - r_{b,j+1} \geq C_j(\alpha_{b,j}) - C_j(\alpha_{b,j+1})
\]

To prevent type \( j+1 \) from misreporting as type \( j \), we require:
\[
r_{b,j+1} - C_{j+1}(\alpha_{b,j+1}) \geq r_{b,j} - C_{j+1}(\alpha_{b,j})
\]
which implies:
\[
r_{b,j+1} - r_{b,j} \geq C_{j+1}(\alpha_{b,j+1}) - C_{j+1}(\alpha_{b,j})
\]
Rewriting:
\[
r_{b,j} - r_{b,j+1} \leq C_{j+1}(\alpha_{b,j+1}) - C_{j+1}(\alpha_{b,j})
\]

Combining the two inequalities:
{\footnotesize
\[
C_j(\alpha_{b,j}) - C_j(\alpha_{b,j+1}) \leq r_{b,j} - r_{b,j+1} \leq C_{j+1}(\alpha_{b,j+1}) - C_{j+1}(\alpha_{b,j})
\]
}
This can be rearranged as:
\[
r_{b,j+1} - C_j(\alpha_{b,j+1}) \leq r_{b,j} - C_j(\alpha_{b,j}) \leq r_{b,j+1} - C_{j+1}(\alpha_{b,j+1})
\]
Therefore, condition (c) must hold.

\subsubsection{Sufficient Conditions}

Now assume that conditions (a), (b), and (c) hold. We will show that the contract \( \mathcal{C} \) satisfies the IR and IC constraints.

\paragraph{Individual Rationality (IR) Constraints}

From condition (a):
\[
U_{b,J} = r_{b,J} - C_J(\alpha_{b,J}) \geq 0
\]
Since \( r_{b,j} \geq r_{b,J} \) and \( C_j(\alpha_{b,j}) \leq C_J(\alpha_{b,J}) \) due to condition (b) and the assumption that costs are non-decreasing in \( \alpha_{b,j} \), we have:
\[
U_{b,j} = r_{b,j} - C_j(\alpha_{b,j}) \geq r_{b,J} - C_J(\alpha_{b,J}) \geq 0
\]
Thus, the IR constraints are satisfied for all \( j \in \mathcal{J} \).

\paragraph{Incentive Compatibility (IC) Constraints}

For any two types \( j \) and \( k \), we need to ensure:
\[
U_{b,j}(\alpha_{b,j}, r_{b,j}) \geq U_{b,j}(\alpha_{b,k}, r_{b,k})
\]

Consider adjacent types \( j \) and \( j+1 \).

For type \( j \), their utility when choosing their own contract:
\[
U_{b,j}(\alpha_{b,j}, r_{b,j}) = r_{b,j} - C_j(\alpha_{b,j})
\]

Utility when misreporting as type \( j+1 \):
\[
U_{b,j}(\alpha_{b,j+1}, r_{b,j+1}) = r_{b,j+1} - C_j(\alpha_{b,j+1})
\]

From condition (c):
\[
r_{b,j} - C_j(\alpha_{b,j}) \geq r_{b,j+1} - C_j(\alpha_{b,j+1})
\]
Thus, type \( j \) prefers their own contract.

Similarly, for type \( j+1 \), their utility when choosing their own contract:
\[
U_{b,j+1}(\alpha_{b,j+1}, r_{b,j+1}) = r_{b,j+1} - C_{j+1}(\alpha_{b,j+1})
\]

Utility when misreporting as type \( j \):
\[
U_{b,j+1}(\alpha_{b,j}, r_{b,j}) = r_{b,j} - C_{j+1}(\alpha_{b,j})
\]

From condition (c):
\[
r_{b,j} - C_{j+1}(\alpha_{b,j}) \leq r_{b,j+1} - C_{j+1}(\alpha_{b,j+1})
\]
Thus, type \( j+1 \) prefers their own contract.

Therefore, the IC constraints are satisfied between adjacent types. By induction, the IC constraints hold for all \( j, k \in \mathcal{J} \).
\end{proof}

\noindent \textit{\textbf{Proof of Lemma~\ref{lemma2}: Optimal Rewards for Given Task Complexities}}

\label{appendix:proofs_lemma2}

\begin{proof}
Given task complexities \( \boldsymbol{\alpha}_b = \{ \alpha_{b,j} \}_{j \in \mathcal{J}} \), we prove that the rewards \( \{ r_{b,j}^* \} \) uniquely minimize the organizer's total cost while satisfying the individual rationality (IR) and incentive compatibility (IC) constraints.

Assume, for contradiction, that there exists another reward set \( \{ \tilde{r}_{b,j} \} \) with a lower total cost:
\[
\sum_{j \in \mathcal{J}} \tilde{r}_{b,j} < \sum_{j \in \mathcal{J}} r_{b,j}^*.
\]
This implies there is some \( j \) such that \( \tilde{r}_{b,j} < r_{b,j}^* \).

For type \( J \), the IR constraint requires:
\[
\tilde{r}_{b,J} \geq C_J(\alpha_{b,J}) = r_{b,J}^*,
\]
so the reduction must occur for some \( j < J \).

Proceeding inductively, assume \( \tilde{r}_{b,j+1} \geq r_{b,j+1}^* \). The IC constraint for type \( j \) is:
\[
\tilde{r}_{b,j} \geq \tilde{r}_{b,j+1} + C_j(\alpha_{b,j}) - C_j(\alpha_{b,j+1}).
\]
Using the definition of \( r_{b,j}^* \):
\[
r_{b,j}^* = C_j(\alpha_{b,j}) + r_{b,j+1}^* - C_j(\alpha_{b,j+1}),
\]
we find that:
\[
\tilde{r}_{b,j} \geq r_{b,j}^*.
\]
This contradicts \( \tilde{r}_{b,j} < r_{b,j}^* \).

Thus, any reward set with a lower total cost violates IR or IC constraints. For uniqueness, suppose another set \( \{ \tilde{r}_{b,j} \} \neq \{ r_{b,j}^* \} \) yields the same total cost. Since \( \tilde{r}_{b,j} \geq r_{b,j}^* \) for all \( j \), equality of total costs implies \( \tilde{r}_{b,j} = r_{b,j}^* \) for all \( j \).
\end{proof}

\noindent \textit{\textbf{Proof of Proposition~\ref{prop1}: Optimal Reward Ratio \( \pi_{b,j}^* \)}}

\label{appendix:proofs_pro1}

\begin{proof}
We aim to determine the optimal reward ratios \( \{ \pi_{b,j}^* \}_{j \in \mathcal{J}} \) that minimize the organizer's total cost while ensuring that the contract \( \mathcal{C} = \{ (\alpha_{b,j}, r_{b,j}) \}_{j \in \mathcal{J}} \) satisfies the Individual Rationality (IR) and Incentive Compatibility (IC) constraints. Since we have already established the feasibility of the contract in Lemma~\ref{lemma1}, we can focus on deriving \( \pi_{b,j}^* \) directly.

From the contract, the reward \( r_{b,j} \) for a bounty hunter of type \( j \) is:
\begin{equation}
r_{b,j} = \pi_{b,j} \cdot p_{b,j} \cdot \sum_{i \in \mathcal{D}_j} R_i
\label{eq:reward_proof}
\end{equation}
where:
\[
p_{b,j} = \sigma(\theta_j - \gamma \alpha_{b,j}) = \frac{1}{1 + e^{-(\theta_j - \gamma \alpha_{b,j})}}
\]

From Lemma~\ref{lemma2}, we have the optimal reward \( r_{b,j}^* \) for each type \( j \). To ensure that the bounty hunters are properly incentivized while minimizing the organizer's cost, we set:
\[
r_{b,j} = r_{b,j}^*
\]
Substituting into Equation~\eqref{eq:reward_proof}:
\[
r_{b,j}^* = \pi_{b,j}^* \cdot p_{b,j} \cdot \sum_{i \in \mathcal{D}_j} R_i
\]
Solving for \( \pi_{b,j}^* \):
\[
\pi_{b,j}^* = \frac{r_{b,j}^*}{p_{b,j} \cdot \sum_{i \in \mathcal{D}_j} R_i}
\]

Substituting the expression for \( p_{b,j} \):
\[
\pi_{b,j}^* = \frac{r_{b,j}^*}{\frac{1}{1 + e^{-(\theta_j - \gamma \alpha_{b,j})}} \cdot \sum_{i \in \mathcal{D}_j} R_i}
\]
Simplifying:
\[
\pi_{b,j}^* = r_{b,j}^* \cdot \left(1 + e^{-(\theta_j - \gamma \alpha_{b,j})}\right) \cdot \left(\frac{1}{\sum_{i \in \mathcal{D}_j} R_i}\right)
\]

\end{proof}

\noindent \textit{\textbf{Optimization of Task Complexities \( \alpha_{b,j}^* \)}}

\label{appendix:proofs_alpha}

\begin{proof}
To determine the optimal task complexities \( \alpha_{b,j}^* \) that minimize the objective function \( \frac{\mathbb{E}[N_u^{(3)}]}{\sum_{j} r_{b,j}^*} \), we proceed as follows.


Assuming the initial expected number of Sybil attackers is \( \mathbb{E}[N_u^{(1)}] \), the expected number after detection by bounty hunters is:
\[
\mathbb{E}[N_u^{(3)}] = \mathbb{E}[N_u^{(1)}] \left( 1 - \frac{1}{M} \sum_{j} p_{b,j} \right)
\]
where \( p_{b,j} = \sigma(\theta_j - \gamma \alpha_{b,j}) = \frac{1}{1 + e^{-(\theta_j - \gamma \alpha_{b,j})}} \) is the detection probability for bounty hunters of type \( j \).

Using the optimal rewards from Lemma~\ref{lemma2}, the total rewards are:
\[
\sum_{j} r_{b,j}^* = \sum_{j} (J - j + 1) c_j \alpha_{b,j}
\]


The organizer's objective function simplifies to:
\[
\min_{\{ \alpha_{b,j} \}} \quad \frac{1 - \frac{1}{M} \sum_{j} p_{b,j}}{\sum_{j} (J - j + 1) c_j \alpha_{b,j}}
\]
Subject to:
\[
\alpha_{b,1} \geq \alpha_{b,2} \geq \dots \geq \alpha_{b,J} \geq 0
\]

\paragraph{Optimization Approach:}

Let \( U = 1 - \frac{1}{M} \sum_{j} p_{b,j} \) and \( V = \sum_{j} (J - j + 1) c_j \alpha_{b,j} \). Our goal is to minimize \( \frac{U}{V} \).

Taking the derivative with respect to \( \alpha_{b,j} \) and setting it to zero:
\[
\frac{\partial}{\partial \alpha_{b,j}} \left( \frac{U}{V} \right) = \frac{ \left( -\frac{1}{M} \frac{\partial p_{b,j}}{\partial \alpha_{b,j}} \right) V - U (J - j + 1) c_j }{ V^2 } = 0
\]
Since \( \frac{\partial p_{b,j}}{\partial \alpha_{b,j}} = -\gamma p_{b,j} (1 - p_{b,j}) \), we have:
\[
\left( \frac{\gamma}{M} p_{b,j} (1 - p_{b,j}) \right) V = U (J - j + 1) c_j
\]
Letting \( \lambda = \frac{U}{V} \), the equation becomes:
\[
\frac{\gamma}{M} p_{b,j} (1 - p_{b,j}) = \lambda (J - j + 1) c_j
\]

\paragraph{Solving for \( p_{b,j} \) and \( \alpha_{b,j} \)}

Define \( a_j = \frac{\lambda M (J - j + 1) c_j}{\gamma} \), then:
\[
p_{b,j} (1 - p_{b,j}) = a_j
\]
Solving the quadratic equation:
\[
p_{b,j} = \frac{1}{2} \left( 1 - \sqrt{1 - 4 a_j} \right)
\]
provided \( a_j \leq \frac{1}{4} \).

Using \( p_{b,j} \), we express \( \alpha_{b,j} \) as:
\[
\alpha_{b,j} = \frac{1}{\gamma} \left( \theta_j + \ln \left( \frac{1 - p_{b,j}}{p_{b,j}} \right) \right)
\]

\paragraph{Determining \( \lambda \)}

Since \( \lambda = \frac{U}{V} \) and both \( U \) and \( V \) depend on \( \lambda \), we solve for \( \lambda \) numerically. The solution must satisfy the ordering constraint on \( \alpha_{b,j} \):
$
\alpha_{b,1} \geq \alpha_{b,2} \geq \dots \geq \alpha_{b,J} \geq 0
$

\end{proof}

\end{document}